\documentclass[twocolumn,envcountsect]{svjour3}          
\smartqed  
\usepackage{graphicx}
\usepackage{mathptmx}      
\usepackage{hyperref}
\usepackage{amssymb}
\usepackage{algorithm}
\usepackage{algorithmic}
\usepackage{mathtools}

\usepackage{xfrac}
\usepackage{changepage}

\usepackage[table]{xcolor}

\definecolor{light-gray}{gray}{0.9}

\usepackage[numbers,sort&compress]{natbib}

\DeclareMathOperator{\sign}{sign}

\journalname{Neuroinformatics}
\begin{document}

\title{Deep Representational Similarity Learning for analyzing neural signatures in task-based fMRI dataset}


\sloppy
\author{Muhammad~Yousefnezhad~\and~Jeffrey~Sawalha~\and~Alessandro~Selvitella~\and~Daoqiang~Zhang}


\institute{M. Yousefnezhad and D. Zhang are with the College of Computer Science and Technology, Nanjing University of Aeronautics and Astronautics, Nanjing 211106, China.\\M. Yousefnezhad and J. Sawalha are with the Deparment of Computing Science and the Deparment of Psychiatry, University of Alberta, Edmonton T6G 2R3, AB, Canada.\\A. Selvitella is with the Department of Mathematical Sciences, Purdue University Fort Wayne, 2101 E Coliseum Blvd, Fort Wayne, 46805 United States.\\\email{\{myousefnezhad, jsawalha\}@ualberta.ca; aselvite@pfw.edu; \\ dqzhang@nuaa.edu.cn\\D Zhang is the corresponding author.}  
}

\date{Received: Feb/2020 / Accepted: Sep/2020}

\maketitle

\begin{abstract}
Similarity analysis is one of the crucial steps in most fMRI studies. Representational Similarity Analysis (RSA) can measure similarities of neural signatures generated by different cognitive states. This paper develops Deep Representational Similarity Learning (DRSL), a deep extension of RSA that is appropriate for analyzing similarities between various cognitive tasks in fMRI datasets with a large number of subjects, and high-dimensionality --- such as whole-brain images. Unlike the previous methods, DRSL is not limited by a linear transformation or a restricted fixed nonlinear kernel function --- such as Gaussian kernel. DRSL utilizes a multi-layer neural network for mapping neural responses to linear space, where this network can implement a customized nonlinear transformation for each subject separately. Furthermore, utilizing a gradient-based optimization in DRSL can significantly reduce runtime of analysis on large datasets because it uses a batch of samples in each iteration rather than all neural responses to find an optimal solution. Empirical studies on multi-subject fMRI datasets with various tasks --- including visual stimuli, decision making, flavor, and working memory --- confirm that the proposed method achieves superior performance to other state-of-the-art RSA algorithms.

\end{abstract}

\section{Introduction}
One of the most significant challenges in both neuroscience and machine learning is comprehending how the human brain works \cite{Kriegeskorte06,Khaligh14}. Indeed, we have long been fascinated by the process of conscious thought, which translates to an interest in better understanding human brains. We anticipate this will offer methods for diagnosing and treating mental health disorders, which could have tremendous benefits \cite{Haxby14}. The neural activities can be analyzed at different levels, but a crucial step is knowing what the similarities (or differences) between distinctive cognitive tasks are \cite{Haxby14,Tony17SDM,Tony17COGN}. It is like a spotlight that allows us to facilitate other areas of brain studies. Since task-based functional Magnetic Resonance Imaging (fMRI) can provide better spatial resolution in comparison with other modalities of measurement, most of the previous studies employed fMRI datasets \cite{Kriegeskorte06,Khaligh14,Haxby14,Tony17SDM,Tony17COGN}.

As one of the fundamental approaches in fMRI analysis, Representational Similarity Analysis (RSA) \cite{Kriegeskorte06,Kriegeskorte08,Connolly12a} evaluates the similarities (or distances) between distinctive cognitive tasks \cite{Walther16}. In practice, classical RSA \cite{Kriegeskorte06,Connolly12a} can be mathematically formulated as a multi-set (group) regression problem --- i.e., a linear model for mapping between the matrix of neural activities and the design matrix \cite{Kriegeskorte06,Cai16}. Classical RSA employs basic linear approaches --- e.g., Ordinary Least Squares (OLS) \cite{Kriegeskorte06} or General Linear Model (GLM) \cite{Connolly12a}.

Recent studies have shown that these methods could not provide accurate performances on large real-world datasets --- such as datasets with a broad Region of Interest (ROI) or whole-brain fMRI data \cite{Khaligh14,Cai16,Tony17NIPS,Khaligh17}. There are several problems in these classical approaches \cite{Walther16,Diedrichsen17}. First, classical RSA methods may not provide accurate similarity analysis --- especially when the covariance of the neural activities has a low rate of Signal-to-Noise Ratio (SNR) \cite{Cai16,Diedrichsen17,Diedrichsen11}. In other words, classical RSA calculates the inverse of the covariance matrix \emph{without a regularization term} \cite{Kriegeskorte06,Kriegeskorte08,Connolly12a,Diedrichsen17}, whereas most fMRI datasets may not be full rank --- i.e., the number of voxels are greater than the time points \cite{Khaligh14,Haxby14,Tony17NIPS}. The next problem is that these techniques need all of the data points in memory at the same time in order to apply the similarity analysis. As a result, they are not computationally efficient for large datasets \cite{Oswal16,Figueiredo16}.

The existing efforts on developing RSA techniques demonstrate some promising results. However, there are still several long-standing challenges. Some of modern approaches focused on the regularization issue \cite{Oswal16,Figueiredo16,Bondell08,Sheng18} --- such as Octagonal Shrinkage and Clustering Algorithm for Regression (OSCAR) \cite{Bondell08} or Ordered Weighted $\ell1$ (OWL) \cite{Oswal16,Figueiredo16}. However, regularized approaches always consider that the relation between the feature space and the design matrix is linear. Alternatively, other algorithms have utilized Bayesian techniques \cite{Cai16,Diedrichsen17,Diedrichsen11,Hans09} --- such as Bayesian RSA (BRSA) \cite{Cai16} or Pattern Component Model (PCM) \cite{Diedrichsen11}. Although Bayesian approaches may significantly handle the SNR issue and even improve some nonlinear datasets, it is limited to a restricted transformation function --- i.e., Gaussian distribution of the hyper-parameter \cite{Tony17NIPS,Sheng18}. Moreover, the time complexity of Bayesian methods is not suitable for high-dimensional datasets because the hyper-parameter must be calculated for each dimension separately \cite{Sheng18}.

This paper proposes Deep Representational Similarity Learning (DRSL) as a new deep extension of the RSA method to solve the previously mentioned challenges. The main contributions of this paper are summarized as follows:
\begin{itemize}
	\item The regular, nonlinear approaches such as the Gaussian kernel suffer from a limited latent space which is caused by a fixed, non-parametric kernel function. Unlike these nonlinear methods, DRSL utilizes a deep network as the parametric kernel function --- including multiple stacked layers of nonlinear transformations. This parametric kernel function allows for a more flexible latent space. Thus, DRSL is not limited to restricted transformation functions.

	\item DRSL also employs a new regularization term that can improve the quality of the results. The proposed regularization function steeply penalizes neural responses with a low rate of SNR.

	\item Most nonlinear similarity approaches (with non-parametric kernels) can significantly increase the computational time of the analysis. Alternatively, DRSL uses a gradient-based optimization method that can generate the neural signatures in each iteration by using a batch of instances instead of utilizing all samples. Consequently, it can provide a time-efficient analysis for evaluating high-dimensional fMRI images  --- such as whole-brain datasets.

	\item We also demonstrate DRSL’s effectiveness by showing that it can distinguish what task the subject is pursuing, even for a novel subject (not in training-set), without a prior calibration phase.

	\item In some sense, the transformation function employed in DRSL is similar to the one utilized in the Deep Canonical Correlation Analysis (DCCA) for multi-view representational learning \cite{Andrew13,Benton17} or Deep Hyperalignment (DHA) for aligning neural activities of multi-subject data across subjects \cite{Tony17NIPS}. However, these supervised classification approaches cannot be used for similarity analysis.
\end{itemize}

The rest of this paper is organized as follows: Section 2 briefly introduces some related works. DRSL is proposed in Section 3. Section 4 reports the empirical studies; Section 5 presents the conclusions and some lines of future work.

\begin{table*}[!h]
	\begin{center}
		\begin{small}
			\rowcolors{1}{light-gray}{white}
			\caption{Variables or Functions}
			\label{tbl:var}
			\begin{tabular}{| c | p{10cm} |}
				\hline
				Variable or Function & Description \\
				\hline
				$\mathbb{R}$ & The set of real numbers.\\
				$i, j, \ell, k, m $ & The indices.\\
				$ T $ & The number of time points.\\
				$ V_{org} $ & The number of voxels in the original space.\\
				$ V $ & The number of mapped features in the linear embedded space.\\
				$ P $ & The number of distinctive categories of stimuli.\\
				$S$ & The number of Subjects.\\
				$ C $ & The number of deep network layers.\\
				$ \mathbf{X}^{(\ell)} = \Big\{ x_{ij}^{(\ell)} \Big\} \in \mathbb{R}^{T \times V_{org}} $ & The original neural activities for $ \ell\text{-}th $ subject.\\
				$ \mathbf{D}^{(\ell)} = \Big\{ d_{ik}^{(\ell)} \Big\} \in \mathbb{R}^{T \times P} $ & The design matrix for $ \ell\text{-}th $ subject.\\
				$ \mathbf{B}^{(\ell)} = \Big\{ \beta_{kj}^{(\ell)} \Big\} \in \mathbb{R}^{P \times V_{org}} $ &  The matrix of estimated regressors for $ \ell\text{-}th $ subject.\\
				$ \widetilde{\mathbf{B}} $ & The mean of the extracted neural signatures across subjects.\\
				$ \mathcal{D}(\widetilde{\beta}_{i.}, \widetilde{\beta}_{j.}) $ & The distance (or similarity) between $ i\text{-}th $ and $ j\text{-}th $ categories of stimuli by using the metric $ \mathcal{D}() $.\\
				$ \theta^{(\ell)} = \big\{ \mathbf{W}_{m}^{(\ell)}, \mathbf{a}_{m}^{(\ell)}, 2\leq m\leq C \big\} $ & All network parameters for $ \ell\text{-}th $ subject.\\
				$\mathbf{h}_{m}^{(\ell)} = \text{g}(\dots)$ & The result of activation function for $\ell\text{-}th$ subject and $m\text{-}th$ layer.\\
				$ U^{(m)}, 2 \leq m < C $ & The number of units in $ m$-$th $ intermediate layer.\\
				$ {\Psi}^{(k,\ell)} $ & A set of randomly selected time points related to $k\text{-}th$ iteration and $\ell\text{-}th$ subject.\\
				$ \alpha \geq 1 $ & The scaling factor.\\
				$ \nabla_{\mathbf{B}}(J_{R}) $ & The gradient of RSA objective function($ J_{R} $) with respect to $ \mathbf{B} $.\\
				$\nabla_{\theta} \big( f(\mathbf{x}_{i.};\theta) \big)$ & The gradient of network parameters.\\ 
				$\eta$ & The learning rate.\\
				$ M $ & The maximum iteration.\\
				$ N $ & The batch size.\\
				$ \mu_{1},\mu_{2},\epsilon $ & The Adam optimization parameters.\\
				$ \delta_{j}, \gamma_{j} $ & The moment vectors.\\
				\hline
				$ r(\mathbf{B}^{(\ell)}) $ & The regularization function.\\
				$ \mathrm{sign}(\mathbf{B}^{(\ell)}) \in \big\{ -1,+1 \big\}^{P\times V} $ & The sign function.\\
				$ g() $ & Nonlinear activation function, i.e., sigmoid, tanh, ReLU.\\
				\hline
			\end{tabular}
		\end{small}
	\end{center}
	\vskip -0.1in
\end{table*}

\section{Background}
In general, a similarity analysis approach maps the neural activities generated based on different cognitive tasks to a set of neural signatures, where each signature belongs to a unique category of stimuli \cite{Haxby14}. Encoding Analysis (EA) \cite{Diedrichsen17,Mitchell08} is utilized for analyzing the various type of neural signatures such as movement stimuli \cite{Sergio05}, low-level visual stimuli \cite{Kay08}, and semantic maps \cite{Mitchell08,Huth16}. The classical encoding approach assumes the neural signatures belong to an individual subject as the basis of a feature space. Here, the neural responses can be characterized as a linear combination of this space \cite{Mitchell08,Naselaris11}.

Representational Similarity Analysis (RSA) is another approach for analyzing similarity across different categories of stimuli \cite{Kriegeskorte06,Haxby14,Diedrichsen17}. Classical RSA can be formulated as a multi-set regression problem, where a set of neural signatures are generated by using Ordinary Least Squares (OLS) \cite{Kriegeskorte06} or General Linear Model (GLM) \cite{Kriegeskorte06,Connolly12a,Cai16}. Recent studies showed that the classical similarity techniques --- such as EA and RSA --- could increase the danger of overfitting mostly on high-dimensional datasets \cite{Cai16,Diedrichsen17}.

Some of the modern RSA methods used regularization functions to control overfitting and improve the performance. Among the prevalent approaches are Ridge Regression \cite{Hoerl70}, Least Absolute Shrinkage and Selection Operator (LASSO) \cite{Sheng18,Yuan06}, Elastic Net \cite{Zou05}, and OSCAR \cite{Bondell08}. Oswal et al. proposed Representational Similarity Learning (RSL) as a modified version of Ordered Weighted $\ell_1$ (OWL) \cite{Figueiredo16} that can produce better performance for analyzing similarity of fMRI datasets in comparison with other sparse techniques --- such as LASSO, Elastic Net, or OSCAR \cite{Oswal16}. Sheng~et~al. also proposed Gradient-Based RSA (GRSA) for applying RSA with LASSO regularization. While GRSA can significantly reduce the time complexity of analysis on the datasets, its performance is not stable. Indeed, the output of GRSA analysis is related to the number of algorithm’s iterations and the learning rate \cite{Sheng18}.

\begin{figure*}[h]
	\centering
	\includegraphics[width=0.95\textwidth]{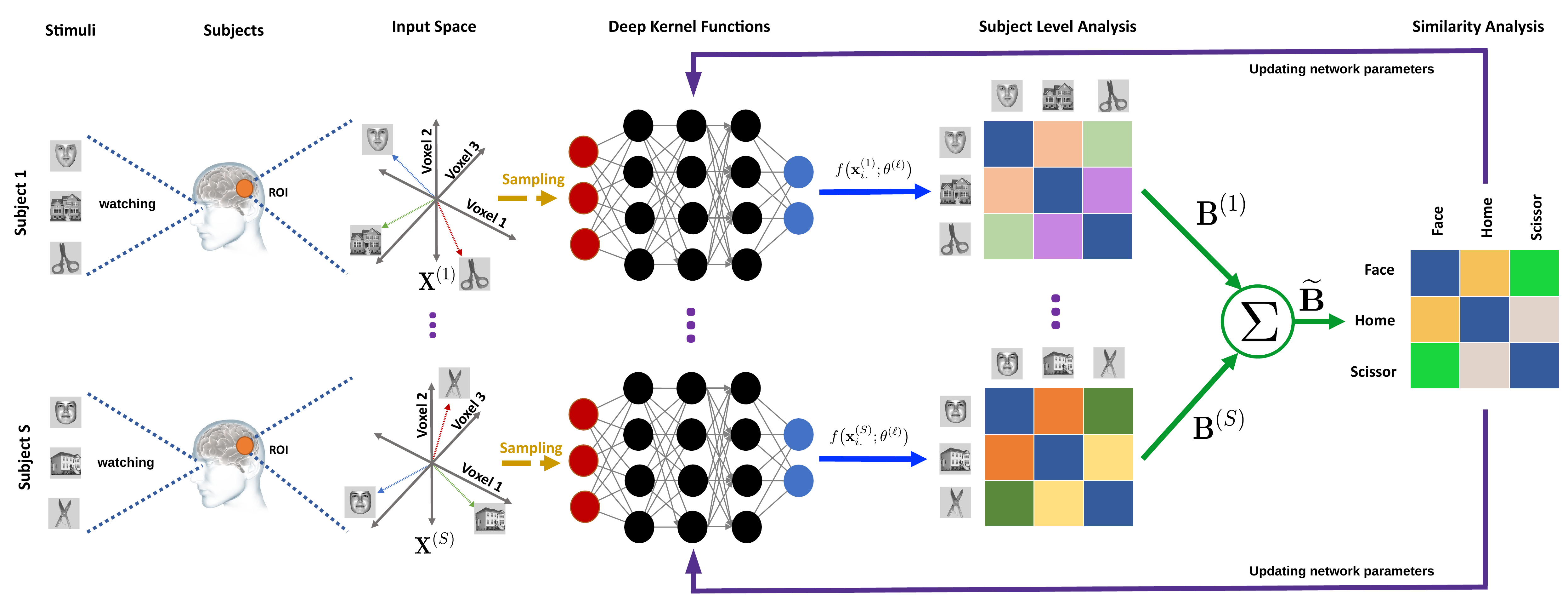}\\
	\caption{The proposed Deep Representational Similarity Learning (DRSL)}
	\label{fig:DRSA}	
\end{figure*}

Similarity analysis for fMRI signals with a low rate of Signal-to-Noise Ratio (SNR) is a challenging problem. Cai~et~al. recently showed that classical RSA could not distinguish the real fMRI signals and randomly generated noises \cite{Cai16}. For solving this issue, they proposed Bayesian RSA (BRSA) considers the covariance matrix as a hyper-parameter generative model \cite{Cai16}. However, the time complexity of BRSA is not suitable for large datasets \cite{Sheng18}. Pattern Component Modeling (PCM) is another Bayesian approach for similarity analysis \cite{Diedrichsen11}. PCM assumes that the neural activities have a multivariate Gaussian distribution, and the model can be evaluated by using the marginal likelihood of the neural responses based on the assumed probability distribution \cite{Diedrichsen17,Diedrichsen11}.

There are two paradigms for evaluating the neural signatures generated by a similarity method, viz, using a distance metric, and converting a similarity analysis to a linear classification problem \cite{Walther16}. Most of the earlier studies in similarity analysis compared the neural signatures by using a distance metric --- such as Euclidean distance, correlation, the rank of generated neural signatures \cite{Kriegeskorte06,Khaligh14,Connolly12a,Walther16}. However, using these metrics for comparing models generated by different similarity approaches sometimes need an extra normalization. For instance, the generated neural signatures must be scaled to a specific range that could be compared by using the Euclidean metric \cite{Walther16,Diedrichsen17}. As another example, the cosine metric can be only utilized to compare two sets of neural signatures (generated by two distinctive similarity techniques) when these signatures have the same origin in the vector space \cite{Walther16}. Alternatively, we can use the correlation metric that naturally provides normalized comparisons for all similarity approaches \cite{Walther16}.

Another paradigm for evaluating the performance of a similarity method is converting the neural signatures into a classification model \cite{Khaligh14,Walther16,Diedrichsen17}. Like a regular classification problem, we first use the subject(s) out cross-validation for partitioning data to training and testing sets \cite{Haxby14}. After that, the neural signatures are generated by applying a similarity approach to the training set \cite{Walther16}. Each pair of these signatures will be used for producing a linear hyperplane as a binary classification model that can distinguish the corresponding categories of stimuli \cite{Walther16}. Further, these binary models can be employed in an Error-Correcting Output Codes (ECOC) framework for predicting multi-class problems \cite{Tony17COGN}. Finally, we can test the generality of the neural signatures by evaluating the performance of the classification models on the testing set, where better signatures can generate more general hyperplane and consequently improve the classification accuracy \cite{Tony17COGN,Walther16}.

\section{Deep Representational Similarity Learning (DRSL)}
Table \ref{tbl:var} illustrates all of the variables and functions that are employed in this paper. We will gradually introduce them in this section. During collecting a task-based fMRI dataset, we generate two elements, viz, functional brain images, and design matrices. fMRI functional image collected from $\ell\text{-}th$ subject can be denoted by $\mathbf{X}^{(\ell)} = \Big\{{x}_{ij}^{(\ell)}\Big\} \in \mathbb{R}^{T\times V_{org}}, 1 \leq i \leq T, 1 \leq j \leq V_{org},$ where $T$ is the number of time points, and $V_{org}$ denotes the number of voxels in the original space. This paper assumes that the neural activities of each subject are column-wise standardized, i.e., $\mathbf{X}^{(\ell)} \sim \mathcal{N}(0,1)$. We can also consider this condition as a preprocessing step if the original data is not standardized. In addition, a design matrix can also be defined for each subject that describes the events related to each cognitive task \cite{Tony17SDM}. In this paper, $\mathbf{D}^{(\ell)} = \Big\{d_{ik}^{(\ell)} \Big\} \in \mathbb{R}^{T\times P}, d_{ik} \in \mathbb{R}, 1 \leq i \leq T, 1 \leq k \leq P$ defines the design matrix belonging to $\ell\text{-}th$ subject. Here, $P$ denotes the number of distinctive categories of stimuli. Further, $\mathbf{d}^{(\ell)}_{.k} \in \mathbb{R}^{T}, 1 \leq k \leq P$ as the $k\text{-}th$ column of the design matrix is the convolution of the onsets of $k\text{-}th$ category ($\mathbf{o}^{(\ell)}_{.k} \in \mathbb{R}^{T}$) with ${\Xi}$ as the Hemodynamic Response Function (HRF) signal ($\mathbf{d}^{(\ell)}_{.k} = \mathbf{o}^{(\ell)}_{.k} * {\Xi}$) \cite{Tony17SDM,Cai16}.

As Figure \ref{fig:DRSA} depicted, DRSL maps neural activities to an information-rich space by using a transformation function, i.e. $\mathbf{x} \in \mathbb{R}^{V_{org}}\to f(\mathbf{x}) \in \mathbb{R}^{V}$, where $V_{org} \geq V$ denotes the number of mapped features in the linear embedded space. Although $f$ can be any restricted fixed transformation function (such as Gaussian or Polynomial), this paper uses multiple stacked layers of nonlinear transformation function in $\mathbf{x}$ as follows:
\begin{equation}
\label{eq:TransFunc}
\begin{gathered}
f\big(\mathbf{x}; \theta\big) = \mathbf{W}_{C}\mathbf{h}_{C-1} + \mathbf{a}_{C},
\end{gathered}
\end{equation}
where the intermediate layers can be defined as follows:
\begin{equation}
\label{eq:Intermid}
\begin{gathered}
\mathbf{h}_{m} = g(\mathbf{W}_{m}\mathbf{h}_{m-1} + \mathbf{a}_{m}),
\end{gathered}
\end{equation}
where $2 \leq m < C$, $\mathbf{h}_{1} = \mathbf{x}$, $C \geq 3$ is the number of deep network layers, $\theta = \big\{\mathbf{W}_{m}, \mathbf{a}_{m} \text{ for } 2 \leq m \leq C \big\}$ denotes all network parameters, and $g$ is a nonlinear function applied componentwise, i.e., Rectified Linear Unit (ReLU), sigmoid, or $tanh$ \cite{Cai16}. By considering $U^{(m)}, 2 \leq m < C$ as the number of units in $m\text{-}th$ intermediate layer, the parameters of the output layer are denoted by $\mathbf{W}_{C} \in \mathbb{R}^{V\times U^{(C-1)}}$, $\mathbf{a}_{C} \in \mathbb{R}^{V}$. For $2 \leq m < C$, the parameters of intermediate layer are defined by $\mathbf{W}_{m} \in \mathbb{R}^{U^{(m)}  \times U^{(m-1)}}$, $\mathbf{a}_{m} \in \mathbb{R}^{U^{(m)}}$ except $\mathbf{W}_{2} \in \mathbb{R}^{U^{(2)}  \times V_{org}}$.

By using the proposed transformation function, the DRSL's objective function can be denoted within subject-level as follows: 
\begin{equation}
\label{eq:DRSL}
\begin{gathered}
J_{R}^{(k,\ell)} = 
\sum_{i \in {\Psi}^{(k,\ell)}}\Big\|f\big(\mathbf{x}_{i.}^{(\ell)}; \theta^{(\ell)}\big) - \mathbf{d}_{i.}^{(\ell)}\mathbf{B}^{(\ell)}\Big\|^2_2   
+ r\Big(\mathbf{B}^{(\ell)}\Big),
\end{gathered}
\end{equation}
where optimal results can be achieved by solving the following optimization problem:
\begin{equation}
\label{eq:DRSLmin}
\begin{gathered}
\underset{\mathbf{B}^{(\ell)}, \theta^{(\ell)}}{\min}\text{ }J_{R}^{(k,\ell)}.
\end{gathered}
\end{equation}
Here, $\mathbf{B}^{(\ell)} = \Big\{ \beta_{kj}^{(\ell)} \Big\} \in \mathbb{R}^{P\times V}, \beta_{kj}^{(\ell)} \in \mathbb{R}, 1 \leq k \leq P, 1 \leq j \leq V$ denotes the matrix of estimated regressors --- including the extracted neural signatures for $\ell\text{-}th$ subject. In addition, ${\Psi}^{(k,\ell)}$ is a set of randomly selected time points related to $k\text{-}th$ iteration and $\ell\text{-}th$ subject. The size of this set ($| {\Psi}^{(k,\ell)} | < T$) is equal to the batch size. Further, $\mathbf{x}_{i.}^{(\ell)} \in \mathbb{R}^{1\times V_{org}}$ denotes all voxels belonging to the $\ell\text{-}th$ subject and $i\text{-}th$ time point (row) of the neural activities $\mathbf{X}^{(\ell)}$, and $\mathbf{d}_{i.}^{(\ell)}\in\mathbb{R}^{1\times P}$ is the $i\text{-}th$ row of the design matrix $\mathbf{D}^{(\ell)}$ that belongs to $\ell\text{-}th$ subject.

Unlike other applications of deep transformation function \cite{Tony17NIPS,Andrew13,Benton17}, we consider a fixed structure of deep network layers for all subjects, including $f(\mathbf{x}_{i.}^{(\ell)}, \theta^{(\ell)})$ rather than $f_{\ell}(\mathbf{x}_{i.}^{(\ell)}, \theta^{(\ell)})$. This notation can improve the stability of generated results and also decrease the number of parameters that must be estimated for each RSA problem. Thus, we just need to estimate additional network parameters ($\theta^{(\ell)}, 1 \leq \ell \leq S$) for each problem in comparison with the classical RSA approaches.

DRSL regularization term is defined as follows:
\begin{equation}
\label{eq:Regularization}
\begin{gathered}
r\Big(\mathbf{B}\Big) =  \sum_{j=1}^{V}\sum_{k=1}^{P} \alpha\Big|\beta_{kj}\Big| + 10\alpha\Big(\beta_{kj}\Big)^2,
\end{gathered}
\end{equation}

where $\alpha$ is the scaling factor. In this paper, we have used $\alpha=10$ for all normalized data --- i.e., $\mathbf{X}^{(\ell)}~\sim~\mathcal{N}(0, 1)$. 


The final neural signatures is also calculated as follows:
\begin{equation}
\label{eq:NeuralSignatures}
\begin{gathered}
\mathbf{\widetilde{B}} = \frac{1}{S}\sum_{\ell=1}^{S} \mathbf{B}^{(\ell)},
\end{gathered}
\end{equation}
where $S$ denotes number of subjects, and $\mathbf{\widetilde{B}}$ calculates the mean of subject-level neural signatures across all subjects. Indeed, $\widetilde{\beta}_{i.} \in \mathbb{R}^{V}\text{, }1 \leq i \leq P$ is the $i\text{-}th$ row of the matrix $\mathbf{\widetilde{B}}$ denotes the extracted neural signatures that belong to $i\text{-}th$ category of stimuli across all subjects. Similarity between two categories of stimuli can be calculated by comparing different rows of the matrix $\mathbf{\widetilde{B}}$. Here, $\mathcal{D}\Big(\widetilde{\beta}_{i.}, \widetilde{\beta}_{j.}\Big)$ denotes the distance (or similarity) between $i\text{-}th \text{ and } j\text{-}th$ categories of stimuli by using the metric $\mathcal{D}()$. We can use any common used metric in machine learning for comparing neural activities --- such as Euclidean, cosine, correlation, covariance, etc. \cite{Khaligh14,Walther16,Khaligh17,Diedrichsen17}. For instance, $\mathbb{E}\bigg[\Big(\mathbf{\widetilde{B}} - \mathbb{E}\big[\mathbf{\widetilde{B}}\big]\Big)\Big(\mathbf{\widetilde{B}} - \mathbb{E}\big[\mathbf{\widetilde{B}}\big]\Big)^\top\bigg]  \in \mathbb{R}^{P \times P}$ denotes the covariance comparison of the neural activities across different categories of stimuli \cite{Cai16}.

\begin{algorithm}[!t]
	\caption{Deep Representational Similarity Learning}
	\label{alg:DRSL}
	\begin{algorithmic}
		\STATE {\bfseries Input:} Dataset $\mathbf{X} = [\mathbf{X}^{(\ell)}\text{, }\ell=1\dots S]$, Design $\mathbf{D} = [\mathbf{D}^{(\ell)}\text{, }\ell=1\dots S]$, Maximum Iteration $M_1$ (default $10$), and Algorithm \ref{alg:DRSLsubject} parameters.
		\STATE {\bfseries Output:}  Neural signatures $\mathbf{\widetilde{B}}$\\
		\STATE {\bfseries Method:}\\
		01. Initializing $\mathbf{\widetilde{B}} \sim \mathcal{N}(0,1)$.\\
		03. \textbf{For} $j = 1\text{:}M_1$\\ 
		04. \quad\textbf{For} $\ell = 1\text{:}S$\\ 
		04. \qquad Generating new $\mathbf{B}^{(\ell)}$ by using Algorithm \ref{alg:DRSLsubject} and $\mathbf{\widetilde{B}}$.\\
		12. \quad\textbf{End For}.\\
		12. \quad Updating $\mathbf{\widetilde{B}}$ by using \eqref{eq:NeuralSignatures}.\\
		13. \textbf{End For}.
	\end{algorithmic}
\end{algorithm}

\subsection{Optimization}
In this section, we implement a version of a Block Coordinate Descent (BCD) optimization approach for the DRSL problem. This method seeks an optimal solution \eqref{eq:DRSL} by using two different steps, which iteratively work in unison. By considering fixed network parameters ($\theta^{(\ell)}$), RSA objective function ($J_R^{(k,\ell)}$) is firstly optimized as follows by using the $k\text{-}th$ mini-batch (${\Psi}^{(k,\ell)}$) of neural activities:
\begin{eqnarray}\label{eq:UpdateBeta}
\begin{gathered}
\underset{\mathbf{B}^{(\ell)}}{\min}\text{ }J_{R}^{(k,\ell)} = \\ 
\underset{\mathbf{B}^{(\ell)}}{\min}\Bigg(
\sum_{i \in {\Psi}^{(k,\ell)}}\Big\|f\big(\mathbf{x}_{i.}^{(\ell)}; \theta^{(\ell)}\big) - \mathbf{d}_{i.}^{(\ell)}\mathbf{B}^{(\ell)}\Big\|^2_2   
+ r\Big(\mathbf{B}^{(\ell)}\Big)\Bigg).
\end{gathered}
\end{eqnarray}
\begin{lemma}\label{lm:GradientB}
	$J_R^{(k,\ell)}$ can be minimized with respect to $\mathbf{B}^{(\ell)}$ as follows:
	\begin{equation}\label{eq:GradientB}
		\begin{gathered}
			\nabla_{\mathbf{B}^{(\ell)}}\Big(J_R^{(k,\ell)}\Big) = \alpha  \sign\Big(\mathbf{B}^{(\ell)}\Big) + 20\alpha \mathbf{B}^{(\ell)} \\ \qquad
			- 2 \sum_{i \in {\Psi}^{(k,\ell)}}\big(\mathbf{d}_{i.}^{(\ell)}\big)^\top\Big(f\big(\mathbf{x}_{i.}^{(\ell)}; \theta^{(\ell)}\big) - \mathbf{d}_{i.}^{(\ell)}\mathbf{B}^{(\ell)}\Big)
		\end{gathered}
	\end{equation}
	where $\sign\Big(\mathbf{B}^{(\ell)}\Big) \in \big\{-1,+1\big\}^{P\times V}$ is the sign function.
	\\Proof. \emph{Please see the proof section.}
\end{lemma}

As the next step, we denote the kernel objective function, where back-propagation algorithm \cite{Rumelhart86} is applied to this function for updating the network parameters:
\begin{equation}\label{eq:Kernel}
\begin{gathered}
\underset{\theta^{(\ell)}}{\min}\sum_{i \in {\Psi}^{(k,\ell)}} 
\Big\|f\big(\mathbf{x}_{i.}^{(\ell)}; \theta^{(\ell)}\big) - \mathbf{d}_{i.}^{(\ell)}\mathbf{B}^{(\ell)}\Big\|^2_2.
\end{gathered}
\end{equation}
In this step, we consider the regressors matrix ($\mathbf{B}^{(\ell)}$) is fixed. The network parameters ($\theta^{(\ell)}$) can be updated by considering the vector $\mathbf{d}_{i.}^{(\ell)}\mathbf{B}^{(\ell)}$ as the ground truth of $f\big(\mathbf{x}_{i.}^{(\ell)}; \theta^{(\ell)}\big)$ and then using back-propagation algorithm to update the parameters. Since $f\big(\mathbf{x}_{i.}^{(\ell)}; \theta^{(\ell)}\big)$ is a standard multilayer perceptron (MLP), we can update the network parameters by using $\nabla_{\theta^{(\ell)}}\Big(f\big(\mathbf{x}_{i.}^{(\ell)}; \theta^{(\ell)}\big)\Big)$, where the output of the optimized deep neural network has the lowest error in comparison with the vector $\mathbf{d}_{i.}^{(\ell)}\mathbf{B}^{(\ell)}$ (as the ground truth). By reducing this error, \eqref{eq:Kernel} will be also minimized. Please refer to \cite{Rumelhart86} for technical information related to MLP and back-propagation algorithm.

Algorithm \eqref{alg:DRSL} shows how DRSL generates the neural signatures from fMRI responses. We first initiate the neural signatures with random numbers --- including $\mathbf{\widetilde{B}} \sim \mathcal{N}(0,1)$. Then, we use Algorithm \eqref{alg:DRSLsubject} for updating these signatures by using the neural responses belonging to each subject. As the first step in Algorithm \eqref{alg:DRSLsubject}, the network parameters are considered fixed, and then Stochastic Gradient Descent (SGD) \cite{Tony17NIPS,Benton17,Kiwiel01} updates the regressors matrix $\mathbf{B}^{(\ell)}$ belonging to $\ell\text{-}th$ subject. As the second step, the regressors matrix is assumed fixed, and then the Adam \cite{Kingma15} algorithm updates the deep network parameters.

This paper develops DRSL as a flexible deep approach for improving the performance of RSA method in fMRI analysis. For seeking an efficient analysis, DRSL uses a deep network (\emph{multiple stacked layers of nonlinear transformation}) for mapping neural activities of each subject into a linear embedded space ($f:\mathbb{R}^{V_{org}}\rightarrow\mathbb{R}^{{V}}$). Unlike the previous nonlinear methods that used a restricted fixed transformation function, mapping functions in DRSL demonstrate flexibility across subjects because they employ multi-layer neural networks, which can implement any nonlinear function \cite{Tony17NIPS,Andrew13,Benton17}. Therefore, DRSL does not suffer from disadvantages of the previous nonlinear approach. Finally, using a gradient-based method to optimize DRSL can provide acceptable time complexity for large datasets because it uses a batch of samples in each iteration rather than all neural responses to find an optimal solution.

\begin{algorithm}[!t]
	\caption{Deep RSL for $\ell\text{-}th$ subject}
	\label{alg:DRSLsubject}
	\begin{algorithmic}
		\STATE {\bfseries Input:} Data $\mathbf{X}^{(\ell)}$, Design $\mathbf{D}^{(\ell)}$, Number of layers $C$, Number of units $U^{(m)}$ for $m=2\text{:}C$, Learning rate $\eta$ (default $10^{-3}$), Maximum Iteration $M_2$ (default $100$), Batch Size $N$ (default $50$), Scaling parameter $\alpha$ (default $10$), Adam optimization parameters $\mu_1 = 0.9$, $\mu_2 = 0.999$, $\epsilon = 10^{-8}$ \cite{Kingma15}, Neural signatures $\mathbf{\widetilde{B}}$.
		\STATE {\bfseries Output:}  Regressors matrix $\mathbf{B}^{(\ell)}$, and Parameters $\theta^{(\ell)}$\\
		\STATE {\bfseries Method:}\\
		01. Initializing $\theta^{(\ell)} \sim \mathcal{N}(0,1)$.\\
		02. Initializing $\mathbf{B}^{(\ell)} := \mathbf{\widetilde{B}}$.\\
		03. $\delta_0 \leftarrow 0$ (Initializing $1^{st}$ moment vector)\\
		04. $\gamma_0 \leftarrow 0$ (Initializing $2^{nd}$ moment vector)\\
		05. \textbf{For} $k = 1\text{:}M_2$\\ 
		06. \quad Creating ${\Psi}^{(k,\ell)}$ by selecting $N$ samples from $1$ to $T$.\\
		07. \quad $\widehat{\phi}_k = \sum_{i \in {\Psi}^{(k,\ell)}}\nabla_{\mathbf{B}^{(\ell)}}\Big(J_R^{(k,\ell)}\Big)$.\\
		08. \quad Updating $\mathbf{B}^{(\ell)} \leftarrow \mathbf{B}^{(\ell)} - \eta\widehat{\phi}_k$.\\
		09. \quad $\phi_k = \sum_{i \in {\Psi}^{(k,\ell)}} \nabla_{\theta^{(\ell)}}\Big(f\big(\mathbf{x}_{i.}^{(\ell)}; \theta^{(\ell)}\big)\Big)$.\\ 
		10. \quad $\delta_k \leftarrow \mu_1\delta_{k-1}+ (1 - \mu_1)\phi_k$.\\
		11. \quad $\gamma_k \leftarrow \mu_2\gamma_{k-1}+ (1 - \mu_2)\phi_k^2$.\\
		12. \quad $\widetilde{\delta}_k \leftarrow \sfrac{\delta_k}{(1 - \mu_1^k)}$.\\
		13. \quad $\widetilde{\gamma}_k \leftarrow \sfrac{\gamma_k}{(1 - \mu_2^k)}$.\\
		14. \quad Updating $\theta^{(\ell)} \leftarrow \theta^{(\ell)} - {\eta\widetilde{\delta}_k} \Big/ {(\sqrt{\widetilde{\gamma}_k} - \epsilon)} $.\\
		15. \textbf{End For}.
	\end{algorithmic}
\end{algorithm}

\begin{table*}[h]
	\caption{The datasets.}
	\vskip 0.02in
	\label{tbl:datasets}
	\begin{small}
		\begin{center}	
			\rowcolors{1}{light-gray}{white}
			\begin{tabular}{llcccccccc}
				\hline
				ID & Title &  Task Type & $S$ & $P$ & $T$ & $V_{ROI}$ & Scanner & TR & TE \\
				\hline
				R105 & Visual object recognition & visual& 6 & 8 & 121 &1452 & G3T & 2500 &30\\
				R107 & Word and object processing& visual &49 & 4 & 164 & 722 & S3T & 2000 &28\\
				R232 & Object-coding localizer task & visual& 10& 4&760  & 9947& S3T& 1060 & 16 \\
				W005 & Mixed-gambles & decision& 16 & 4 & 714 & -- & S3T & 2000 & 30  \\
				W203 & Visual imagery and false memory& memory& 26 &4 &534 & -- & S1.5T & 2000 & 35 \\
				W231 & Integration of sweet taste & flavor& 9 & 6 & 1119 & -- & S3T & 2000  & 30\\
				\hline
			\end{tabular}
		\end{center}
		S is the number of subject; P denotes the number of stimulus categories; T is the number of scans in unites of TRs (Time of Repetition); $V_{ROI}$ denotes the number of voxels in ROI;  
		$19742$ voxels are extracted from \emph{MNI152-T1-4mm} space \cite{Kriegeskorte06} for all whole-brain datasets. Scanners include S=Siemens, or G = General Electric in 1.5, or 3 Tesla; TR is Time of Repetition in millisecond; TE denotes Echo Time in millisecond; Please see \url{https://openneuro.org/} for more information.
	\end{small}
\vskip -0.1in
\end{table*}

\section{Experiments}
\subsection{Datasets}
Table \ref{tbl:datasets} shows the 6 datasets that are employed in this paper for running empirical studies. These datasets are shared by Open~Neuro\footnote{Available at \url{https://openneuro.org/}}. In this paper, there are two groups of datasets --- including ROI-based data, and the whole-brain data. We employ a prefix `R' in the rest of this paper for denoting ROI datasets --- i.e., R105, R107, and R232. Further, a prefix `W' will be used to define the whole-brain datasets --- such as W005, W203, and W231. These datasets are listed as follows:

\begin{itemize}
\item \textbf{R105}: visual object recognition includes eight different categories of visual stimuli --- including grayscaled photos of faces, houses, cats, bottles, scissors, shoes, chairs, and scramble patterns. The neural activities in temporal cortex (VT) is selected as the ROI for this dataset \cite{Haxby14,Tony17COGN}.
\item \textbf{R107}: word and object processing contains four categories of stimuli --- i.e., words, consonants, objects, and scramble photos. ROI in this dataset is selected based on the original paper \cite{Duncan09}.
\item \textbf{R232}: object-coding localizer task includes four tasks --- including faces, objects, places, and scramble photos. We have used the original ROI that is presented in the main reference of this dataset \cite{Carlin17}.
\item \textbf{W005}: mixed-gambles is the first whole-brain data that contains four decision making tasks. Each cognitive tasks in this dataset has the same change of selection (25\%). The scheme of experiment is explained in the original paper \cite{Tom07}.
\item \textbf{W203}: visual imagery and false memory contains four categories of working memory tasks --- including hit, omission, false alarm, and correct rejection. During the experiment, two groups of stimuli presented to each subject in random order --- i.e., 45 words and 45. In the recall phase, subjects need to remember whether a picture of the item had
been presented, or only a word. Please refer to the main reference for more information \cite{Stephan17}.
\item \textbf{W231}: integration of sweet taste includes six different categories of flavor task --- including 0 cal, 112 cal, 150 cal, rinse, tasteless, and control. The setup of experiment is presented in \cite{Veldhuizen17}.
\end{itemize}

All datasets are separately preprocessed by FSL 6.0.1\footnote{Availabel at \url{https://fsl.fmrib.ox.ac.uk}} --- i.e., slice timing, anatomical alignment, normalization, smoothing. For the whole-brain datasets, we have registered all fMRI images to the \emph{MNI152-T1-4mm} standard space, and then $V_{org}=19742$ voxels are extracted from the standard whole-brain mask [4]. We have also provided a preprocessed version of these datasets in MATLAB format\footnote{Available at \url{https://easydata.gitlab.io}}.

\subsection{Performance analysis}
We compared the performance of DRSL with 6 existing similarity methods. The hyper-parameters for each similarity technique are selected based on a grid-search. Here, we select the hyper-parameters for each method that generate a lower error for the corresponding objective function. The number of maximum iterations for all algorithms is considered $1000$. For the DRSL method, we have also selected $10$ iterations for Algorithm~\ref{alg:DRSL} and $100$ iterations for Algorithm~\ref{alg:DRSLsubject} --- i.e., the total iterations are $10\times 100 = 1000$. As a non-parametric approach, this paper employs classical RSA with GLM optimization as a baseline \cite{Kriegeskorte06,Connolly12a}. We also report the performance of the LASSO algorithm \cite{Yuan06}, where different values for balance factor $\alpha_{LASSO} = [0, 0.8, 0.9, 1, 1.1]$ are evaluated, and $\alpha_{LASSO} = 0.9$ generates the best results. As another regularized method, the performance of RSL is presented. To generate the results, we evaluate the performance of the RSL approach by using two regularization terms proposed in the original paper \cite{Oswal16} --- i.e., GrOWL-Lin and GrOWL-Spike. Here, GrOWL-Spike produces a better performance for similarity analysis. 

The performance of Bayesian RSA (BRSA) \cite{Cai16,Cai19} and Pattern Component Modeling (PCM) \cite{Diedrichsen17,Diedrichsen11} are reported here. We evaluate the performance of these techniques by using different nuisance regressor methods --- including Principal Component Analysis (PCA), Independent Component Analysis (ICA), Factor Analysis, and Sparse PCA. In all cases, PCA generates better performance and runtime. We also use different SNR prior --- i.e., Gaussian, Log-Norm, and Uniform. The Gaussian prior generates better results. We also initiate $20$ as the minimum number of iterations and consider a full rank of the covariance matrices in these approaches. We also compare the performance with Gradient RSA (GRSA) \cite{Sheng18} --- by using $\alpha_{GRSA}=[0, 0.8, 0.9, 1]$, the learning rates $[10^{-2}, 10^{-3}, 10^{-4}]$, and the batch sizes $[10, 30, 50, 80, 100]$. The best results are generated by using $\alpha_{GRSA}=0.9$, the learning rate $10^{-3}$, with a batch size of $50$.

This paper also reports the performance of the proposed method using linear and deep transformation functions. Linear RSL (LRSL) utilizes the objective function \eqref{eq:DRSL} and the optimization Algorithm \ref{alg:DRSL}, but the transformation function is considered linear --- i.e., $f\big(\mathbf{x}\big)=\mathbf{x}$. Here, we illustrate how much DRSL improves the performance of similarity analysis with the proposed deep transformation function. We have generated results in both LRSL and DRSL by using different values of $\alpha = [1, 5, 10, 20, 50, 100]$. In all datasets with normalization $\mathbf{X}^{(\ell)} \sim \mathcal{N}(0,1)$,  $\alpha = 10$ produces a better performance in comparison with other values. Further, the Adam optimization parameters are set optimally based on the original paper \cite{Kingma15}, including $\mu_1=0.9$, $\mu_2=0.999$, and $\epsilon=10^{-8}$. We also evaluate the performance of LRSL and DRSL by using different learning rates $[10^{-2}, 10^{-3}, 10^{-4}]$ and batch sizes $[10, 30, 50, 80, 100]$. Like GRSA, the best results are produced by using the batch size $50$ and learning rate $10^{-3}$.

\begin{table*}[!h]
	\caption{Between-class correlation analysis (max$\pm$std)}
	\vskip -0.15in
	\label{tbl:Cor}
	\begin{adjustwidth}{0cm}{}
		\begin{center}\begin{small}
				\rowcolors{1}{white}{light-gray}	
				\begin{tabular}{lcccccccc}
					\hline
					Datasets & RSA \cite{Kriegeskorte08}  & LASSO \cite{Oswal16,Yuan06} & RSL \cite{Oswal16} & BRSA \cite{Cai16} & PCM \cite{Diedrichsen11} & GRSA \cite{Sheng18} & LRSL & DRSL \\
					\hline
					R105 & 0.947$\pm$0.042 &  0.751$\pm$0.242 & 0.821$\pm$0.120 & 0.389$\pm$0.010 & 0.401$\pm$0.064 & 0.584$\pm$0.142 &0.451$\pm$0.081 & \textbf{0.372$\pm$0.016} \\
					R107 & 0.922$\pm$0.053 &  0.715$\pm$0.147 & 0.531$\pm$0.123 & 0.458$\pm$0.076 & 0.357$\pm$0.041 & 0.426$\pm$0.099 & 0.142$\pm$0.092 & \textbf{0.135$\pm$0.000} \\
					R232 & 0.927$\pm$0.033 &  0.900$\pm$0.026 & 0.631$\pm$0.193 & 0.871$\pm$0.100 & \textbf{0.490$\pm$0.057} & 0.754$\pm$0.107 & 0.555$\pm$0.112 & 0.496$\pm$0.093 \\
					W005 & 0.891$\pm$0.035 &  0.823$\pm$0.036 & 0.699$\pm$0.076 & 0.519$\pm$0.045 & 0.591$\pm$0.083 & 0.392$\pm$0.069 & 0.361$\pm$0.021 & \textbf{0.139$\pm$0.019} \\
					W203 & 0.951$\pm$0.045 &  0.851$\pm$0.077 & 0.572$\pm$0.155 & 0.383$\pm$0.051 & 0.405$\pm$0.075 & 0.411$\pm$0.094 & 0.451$\pm$0.012 & \textbf{0.270$\pm$0.039} \\
					W231 & 0.888$\pm$0.113 &  0.761$\pm$0.114 & 0.421$\pm$0.273 & 0.521$\pm$0.082 & 0.487$\pm$0.107 & 0.267$\pm$0.243 & 0.231$\pm$0.063 & \textbf{0.126$\pm$0.002} \\
					\hline
	\end{tabular}\end{small}\end{center}\end{adjustwidth}
	\vskip -0.15in
\end{table*}
\begin{table*}[!h]
	\caption{Accuracy of classification analysis (mean$\pm$std)}
	\vskip -0.05in
	\label{tbl:Cls}
	\begin{adjustwidth}{0cm}{}
		\begin{center}\begin{small}
				\rowcolors{1}{white}{light-gray}	
				\begin{tabular}{lcccccccc}
					\hline
					Datasets & RSA \cite{Kriegeskorte08}  & LASSO \cite{Oswal16,Yuan06} & RSL \cite{Oswal16} & BRSA \cite{Cai16} & PCM \cite{Diedrichsen11} & GRSA \cite{Sheng18} & LRSL & DRSL \\
					\hline
					R105 & 18.69$\pm$2.37 &  47.63$\pm$1.04 & 51.29$\pm$1.13 & 60.39$\pm$0.69 & 63.91$\pm$0.92 & 58.36$\pm$1.03 &59.73$\pm$2.81 & \textbf{78.13$\pm$0.19} \\
					R107 & 33.57$\pm$3.06 &  38.18$\pm$1.27 & 37.97$\pm$2.08 & 65.24$\pm$0.58 & 69.11$\pm$0.73 & 66.05$\pm$1.76 & 72.31$\pm$0.04 & \textbf{84.26$\pm$0.64} \\
					R232 & 31.78$\pm$2.71 &  43.66$\pm$1.80 & 50.30$\pm$1.27 & 61.06$\pm$0.19 & 59.74$\pm$0.88 & 65.14$\pm$1.01 & 63.27$\pm$1.30 & \textbf{70.31$\pm$0.55} \\
					W005 & 27.10$\pm$1.02 &  31.03$\pm$0.71 & 46.17$\pm$1.36 & 66.58$\pm$0.73 & 68.24$\pm$0.93 & 61.43$\pm$0.92 & 60.00$\pm$0.72 & \textbf{81.37$\pm$0.11} \\
					W203 & 24.40$\pm$2.13 &  35.64$\pm$0.81 & 60.06$\pm$2.41 & 76.90$\pm$0.59 & 72.58$\pm$0.33 & 83.15$\pm$0.24 & 75.68$\pm$0.80 & \textbf{91.40$\pm$0.24} \\
					W231 & 18.63$\pm$3.37 &  29.43$\pm$0.96 & \textbf{68.25$\pm$1.07} & 50.26$\pm$1.00 & 49.51$\pm$0.98 & 52.76$\pm$1.61 & 51.68$\pm$1.42 & 65.74$\pm$1.00 \\
					\hline
	\end{tabular}\end{small}\end{center}\end{adjustwidth}
	\vskip -0.2in
\end{table*}

In DRSL, we need to define the structure of deep neural networks --- i.e., multiple stacked layers of nonlinear transformations. Since we want to use the deep network as the kernel function in a multi-set regression problem, the network structure must satisfy some criteria --- i.e., number of layers, number of units, and the nonlinear activation functions. These criteria can affect both the performance and runtime of the analysis. As the previous studies showed \cite{Andrew13,Benton17,Tony17NIPS}, deep networks with more than two layers can act as nonlinear kernel functions for accurately modeling complex real-world datasets. Since the number of samples is limited in most fMRI datasets, we keep the number of layers in the deep network minimal \cite{Tony17NIPS}. Thus, we have employed two hidden layers for DRSL --- i.e., $C = 4$. Another component of the deep network is the number of units. Here, we present the number of units for the hidden layers and the output layer as follows, $[HL1, HL2, OUT]$. As an example, $[1000, 700, 500]$ represents $U^{(1)}= 1000$ as the number of the first hidden layer, $U^{(2)}=700$ as the number of the second hidden layer, and $V=500$ as the output layer. We evaluate the DRSL by using different numbers of units in each layer --- including $[2000, 1500, 1000, 700, 500, 200, 100, 50]$. The best results for all datasets except $R107$ are achieved by using  $[1000, 700, 500]$ settings. For $R107$, a network with $[700, 500, 200]$ setting is generated the best results (perhaps because the number of voxels in this data is less than $1000$). The last component in the network structure is the nonlinear activation function. Since the deep network is used in the form of a multi-set regression problem, the output layer must have a linear activation function \cite{Andrew13,Benton17,Tony17NIPS}. So, we only need to define the activation functions for the hidden layers. We evaluated the performance of DRSL by employing different nonlinear activation functions --- including ReLU, sigmoid, $tanh$. In most of the normalized datasets, the sigmoid activation function generated better performance. Based on our empirical studies, the explained structure for the DRSL method was fit to our hardware limitation and provided an efficient trade-off between runtime and performance (please see the result sections).

All algorithms in this paper are implemented by Python $3$ and run on a PC with certain specifications\footnote{Main:~Giga X399, CPU:~AMD Ryzen Threadripper 2920X~(24$\times$3.5~GHz), RAM:~64GB, GPU:~NVIDIA GeForce RTX 2080 SUPER~(8GB memory), OS:~Fedora~31, Python:~3.7.5, Pip:~19.3.1, Numpy:~1.16.5, Scipy:~1.2.1, Scikit-Learn:~0.21.3, MPI4py:~3.0.1, PyTorch:~1.2.0.} by authors for generating the experimental results. The source codes of the proposed method and a GUI-based toolbox is also shared\footnote{Available at \url{https://easyfmri.gitlab.io}}.

\begin{figure*}[!t]
	\begin{center}
		\begin{minipage}{0.7\linewidth}
			\begin{minipage}{0.32\linewidth}
				\includegraphics[width=0.98\textwidth,height=\linewidth]{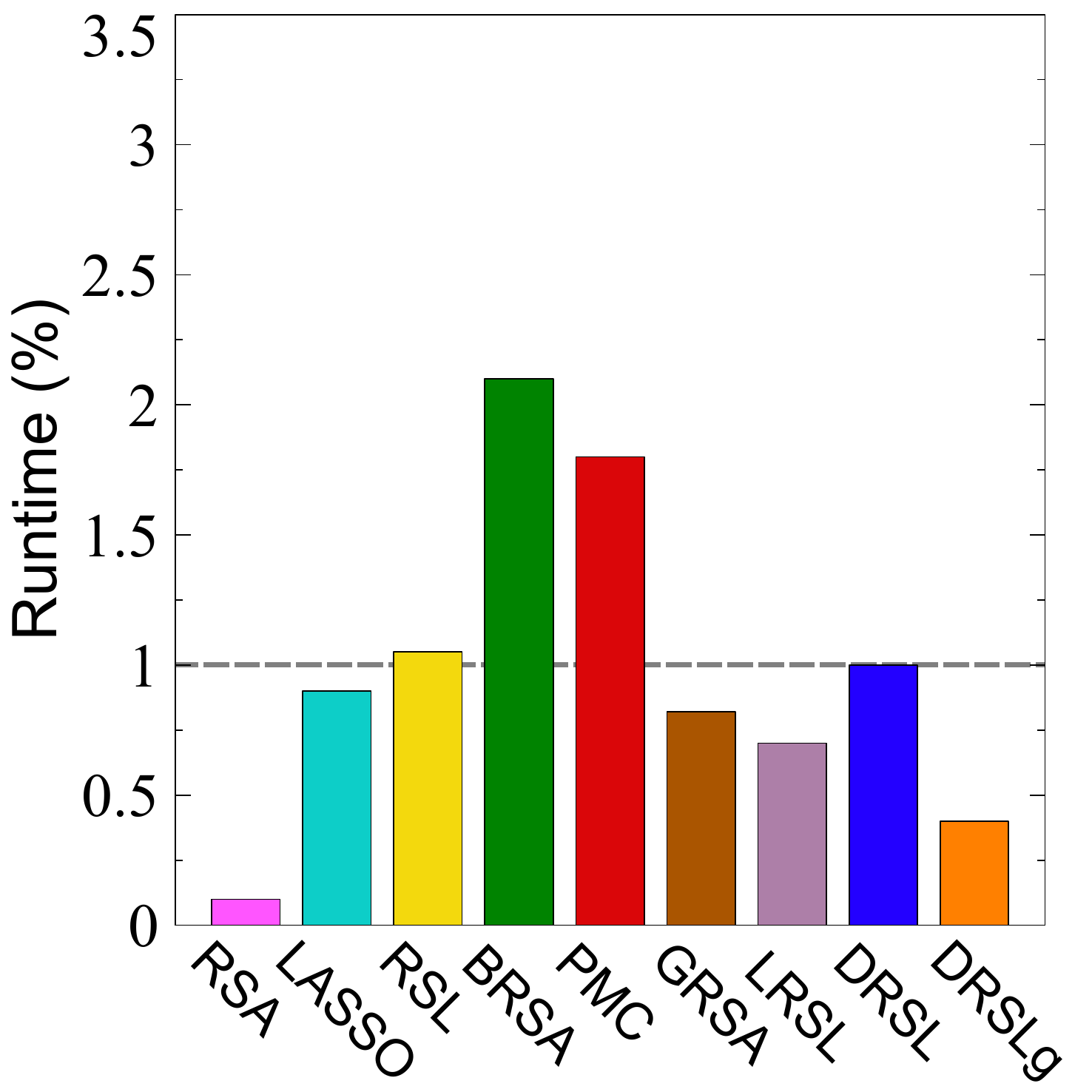}\\
				\centering {\small (a) R105}
			\end{minipage}
			\begin{minipage}{0.32\linewidth}
				\includegraphics[width=0.98\textwidth,height=\linewidth]{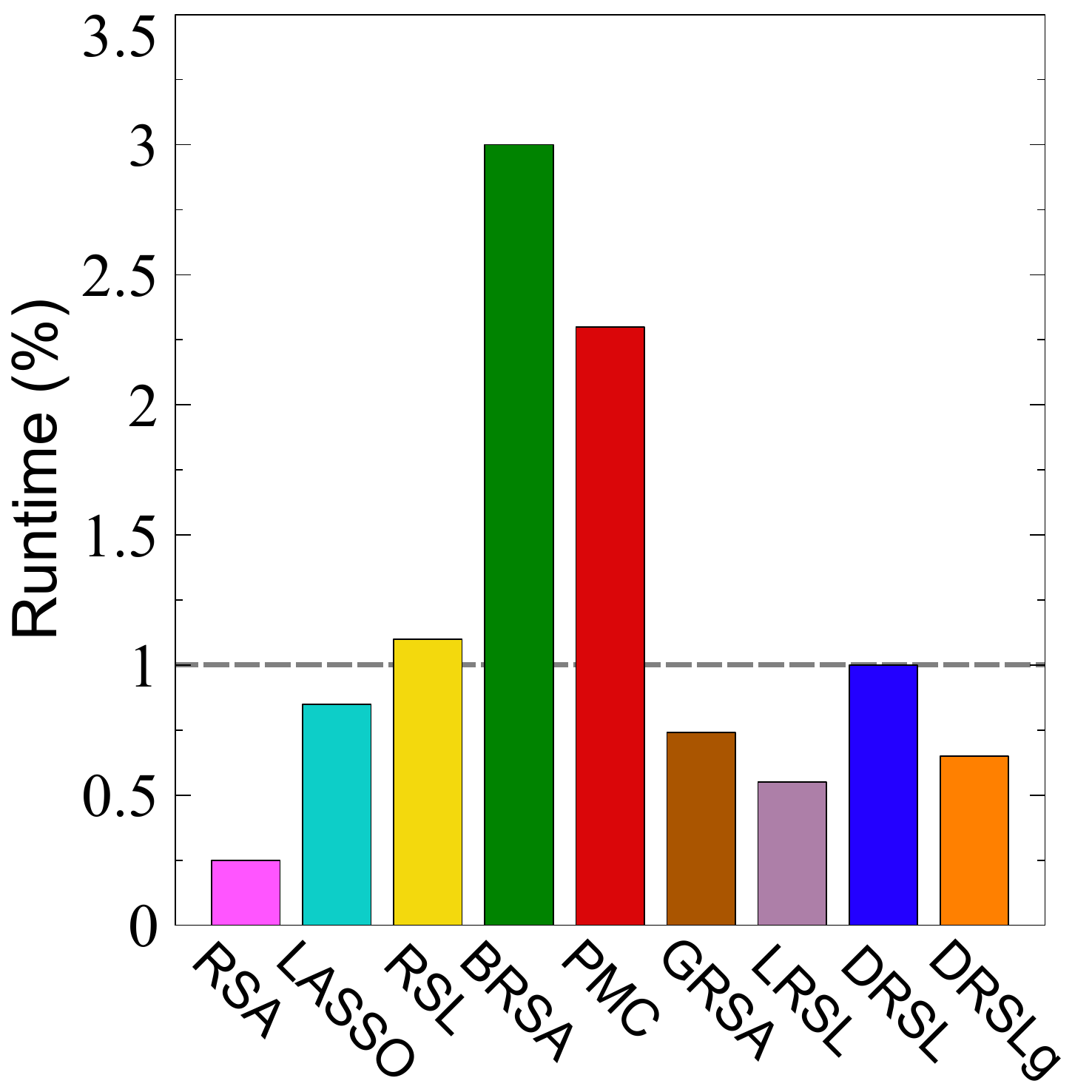}\\
				\centering {\small (b) R107}
			\end{minipage}
			\begin{minipage}{0.32\linewidth}
				\includegraphics[width=0.98\textwidth,height=\linewidth]{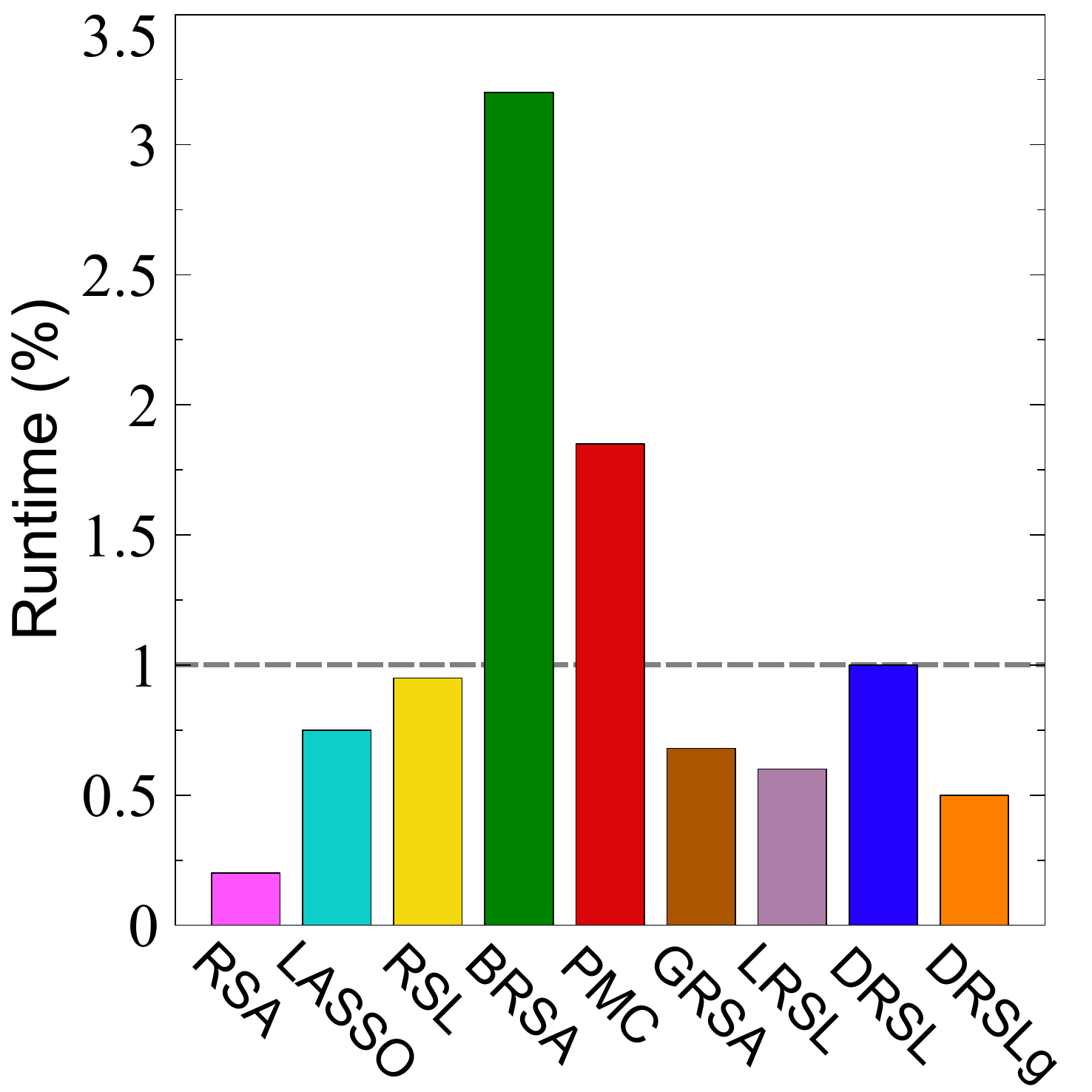}\\
				\centering {\small (c) R232}
			\end{minipage}
			\caption{Runtime Analysis}
			\label{fig:Runtime}
		\end{minipage}		
		\begin{minipage}{0.29\linewidth}
			\begin{minipage}{0.99\linewidth}
				\includegraphics[width=0.8\textwidth]{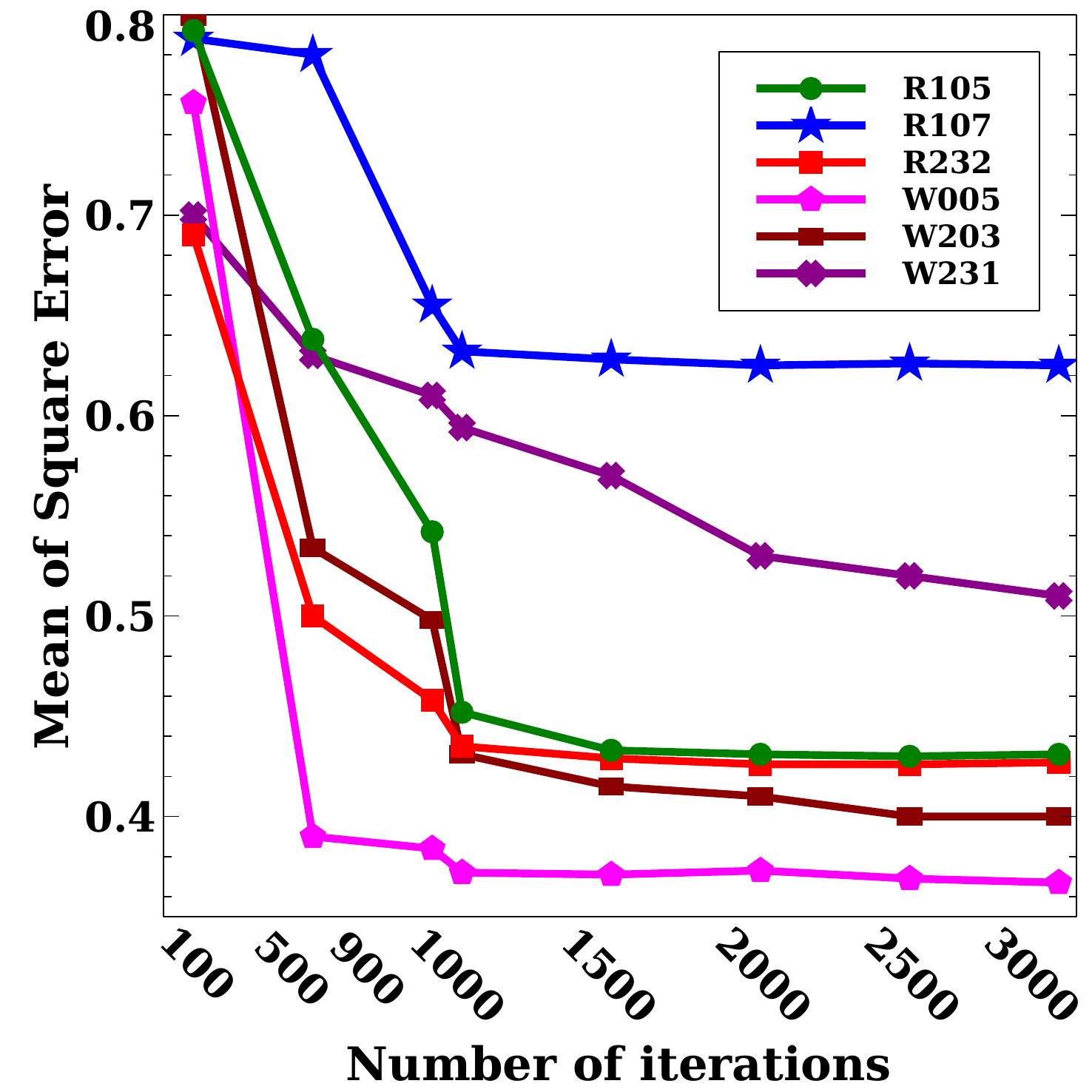}
			\end{minipage}
			\caption{Analyzing different number of iterations in DRSL}
			\label{fig:Iter}	
		\end{minipage}			
	\end{center}
	\vskip -0.1in
\end{figure*}

\begin{figure*}[!h]
	\begin{center}
		\begin{minipage}{0.48\linewidth}
			\includegraphics[width=0.9\textwidth]{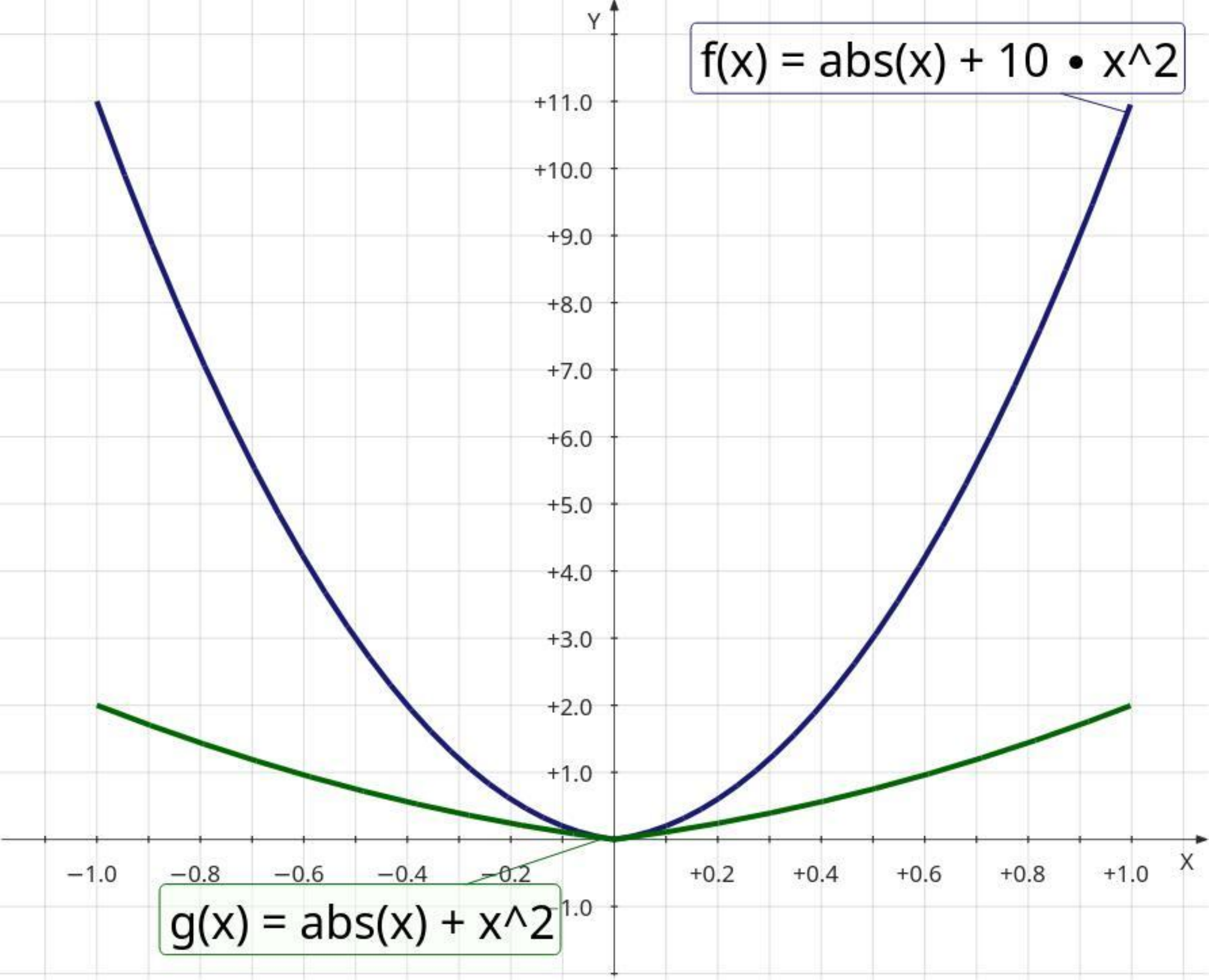}\\
			\centering (A) $\alpha = 1$
		\end{minipage}
		\begin{minipage}{0.48\linewidth}
			\includegraphics[width=0.9\textwidth]{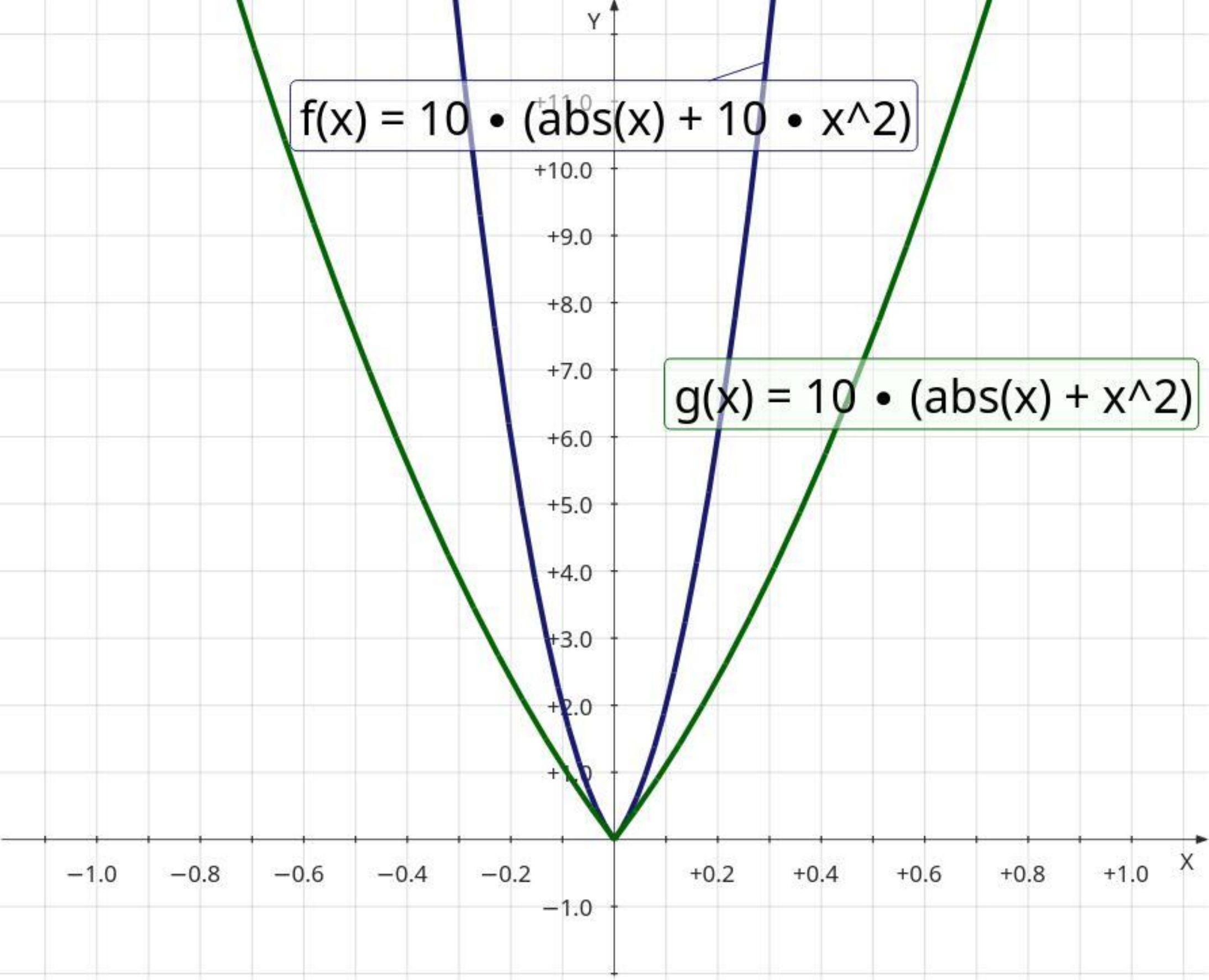}\\
			\centering (B) $\alpha = 10$
		\end{minipage}
		\caption{Regularization function $r(\mathbf{B})$ by using $\alpha = 1$ and $\alpha = 10$.}
		\label{fig:RT}	
	\end{center}
	\vskip -0.3in
\end{figure*}

\subsubsection{Correlation analysis}
In this section, we use the mentioned methods for generating the neural signatures. Then, we used two performance tests for evaluating the quality of the generated signatures --- including correlation and classification analysis. Suppose $\mathbf{\widetilde{B}} = \big\{ \widetilde{b}_{ij} \big\} \in \mathbb{R}^{P\times V}$ denotes the generated neural signatures by using each of similarity approaches for $P$ different categories of stimuli. Further, $\mathbf{\widetilde{b}}_{i.} \in \mathbb{R}^{V}$ represents the signature belonging to $i\text{-}th$ category of stimuli. The primary assumption for similarity analysis is that better methods must generate independent neural signatures in distinctive categories \cite{Kriegeskorte06,Khaligh14,Connolly12a,Walther16,Khaligh17,Diedrichsen17}. Thus, this paper analyzed the similarities of extracted signatures in different categories of stimuli by correlation metric. For evaluating the performance of each similarity technique, we first analyze between-class correlation --- i.e., ${\rho} (\mathbf{\widetilde{B}}) = \underset{i=1}{\overset{P}{\max}}\underset{j=i+1}{\overset{P}{\max}} | \text{corr}(\mathbf{\widetilde{b}}_{i.}, \mathbf{\widetilde{b}}_{j.}) |$. Here, a better similarity approach must generate more independent neural signatures and reduce the correlation among absolute values simultaneously. As mentioned before, correlation is a normalized metric that can be used for evaluating signatures generated by using different similarity approaches \cite{Walther16}.

Table \ref{tbl:Cor} shows the maximum of the between-class correlation. The neural signatures generated by classical RSA are highly correlated. LASSO and RSL provided better signatures by applying regularization approaches. Here, RSL outperformed LASSO by using a customized regularization term that can handle sparsity and noise in comparison with the regular $\ell_1$ norm. Next, BRSA and PCM generated neural signatures by estimating a Gaussian distribution on the fMRI datasets. While PCM could outperform other algorithms in R232, the Gaussian prior assumption cannot provide better performance in comparison with the proposed approach. As mentioned before, GRSA cannot provide stable analysis. Table \ref{tbl:Cor} compares the variance of the neural signatures generated by GRSA with other techniques. Finally, DRSL has produced better performance in comparison with other methods. Indeed, it provides better feature representation by using deep neural networks, i.e., mapping the neural activities to a linear space.

\subsubsection{Classification analysis}
We first use one-subject-out cross-validation for partitioning each dataset into the training and testing sets. Then, training-set is used to generate the neural signatures ($\mathbf{\widetilde{B}} \in \mathbb{R}^{P\times V}$). As mentioned before, each row of these signatures belong to a unique category of stimuli. The goal is to generate a linear hyperplane by using each pair of these signatures where this hyperplane can be used as a binary classification model to distinguish two corresponding categories of stimuli. Suppose $\mathbf{a}_{ij}\mathbf{\widehat{x}} + {z}_{ij} = 0$ represents a linear hyperplane for classifying between $i\text{-}th$ and $jk\text{-}th$ classes, where $\mathbf{a}_{ij} = \Big(\mathbf{\widetilde{b}}_{i.} - \mathbf{\widetilde{b}}_{j.}\Big)\Sigma_{ij}$. Here, $\Sigma_{ij}$ is used for normalization and denotes the square root of the residuals error --- i.e., please refer to equation (4) in \cite{Walther16}. Further, $\mathbf{\widehat{x}} \in \mathbb{R}^{V}$ denotes a neural response and ${z}_{ij} \in \mathbb{R}$ is the offset from the origin of the vector space, which its optimal value calculates by applying the training-set to the hyperplane \cite{Walther16,Diedrichsen17}. After generating the binary classifiers, we use these models in an ECOC framework as a multi-class approach, where labels for testing-set are assigned to the closest Hamming distance \cite{Tony17COGN}. To summarize, we first generate the neural signatures and then the binary and multi-class models by using the training-set. After that, we use the testing-set to evaluate the accuracy of these models. It is worth noting that the testing-set is entirely isolated from training-phase. In addition, the DRSL's network parameter for testing-set is estimated only by using the design matrices in the testing data --- i.e., we use \eqref{eq:DRSL} in testing-phase when only testing-set is employed for calculating network parameters.

Table \ref{tbl:Cls}  illustrates the classification analysis. In this analysis, the performance of RSA (without regularization term) is akin to random sampling. The results of LASSO and RSL showed the positive effect of regularization on the performance of the linear models. By fitting a Gaussian distribution to the data manifold, BRSA and PCM provided better performance in comparison with the method that only control the quality of results via a regularization function. By comparing the linear LRSL with the proposed DRSL, we can understand the positive effect of the nonlinear transformation function on the performance of the analysis. DRSL provides better performance in most of the fMRI datasets because of two critical components: 1) a flexible nonlinear kernel, 2) an effective regularization approach.

\begin{figure*}[!h]
	\begin{center}
		\begin{minipage}{0.22\linewidth}
			\includegraphics[width=0.98\textwidth]{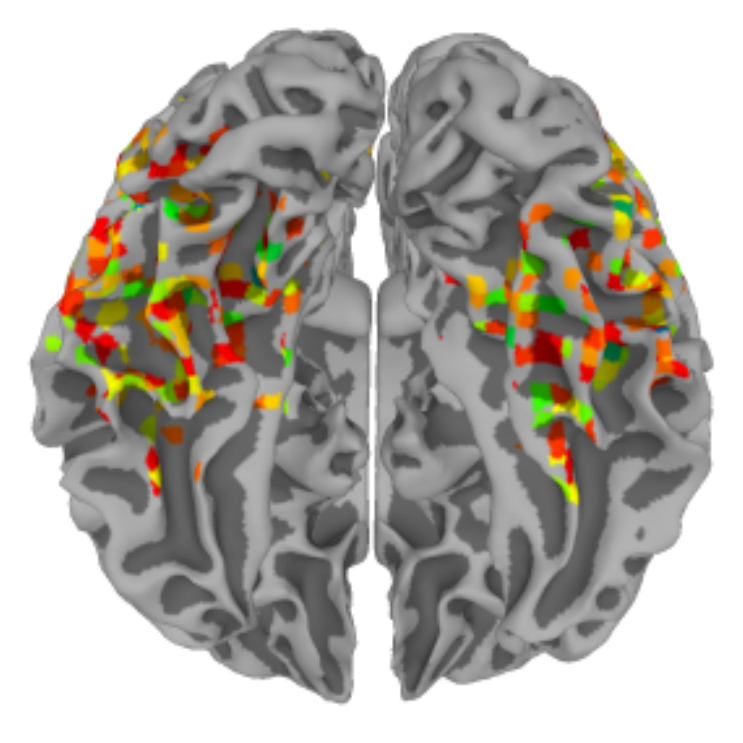}\\ 
			\centering (A) R105--Scissor
		\end{minipage}
		\begin{minipage}{0.22\linewidth}
			\includegraphics[width=0.98\textwidth]{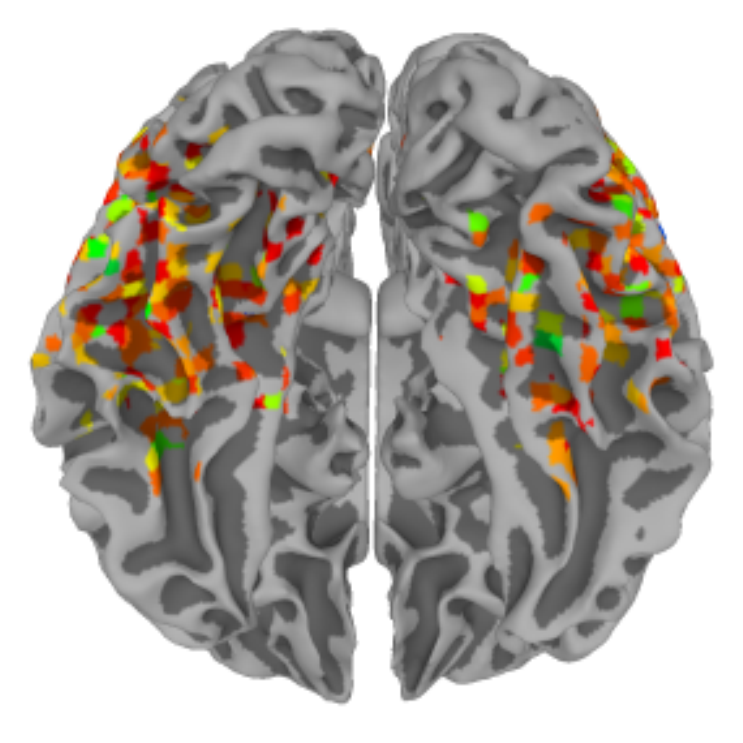}\\
			\centering (B) R105--Face
		\end{minipage}
		\begin{minipage}{0.22\linewidth}
			\includegraphics[width=0.98\textwidth]{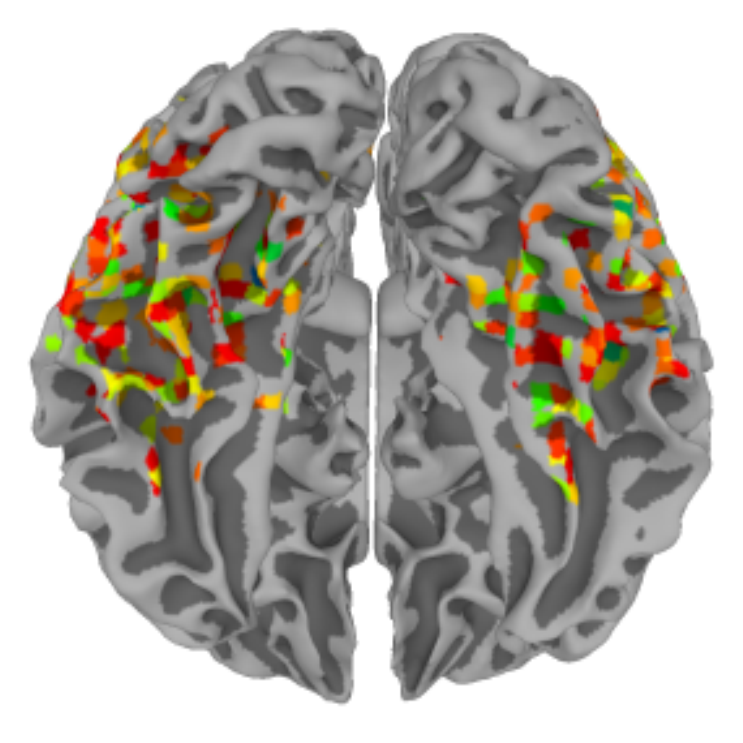}\\
			\centering (C) R105--Cat
		\end{minipage}
		\begin{minipage}{0.22\linewidth}
			\includegraphics[width=0.98\textwidth]{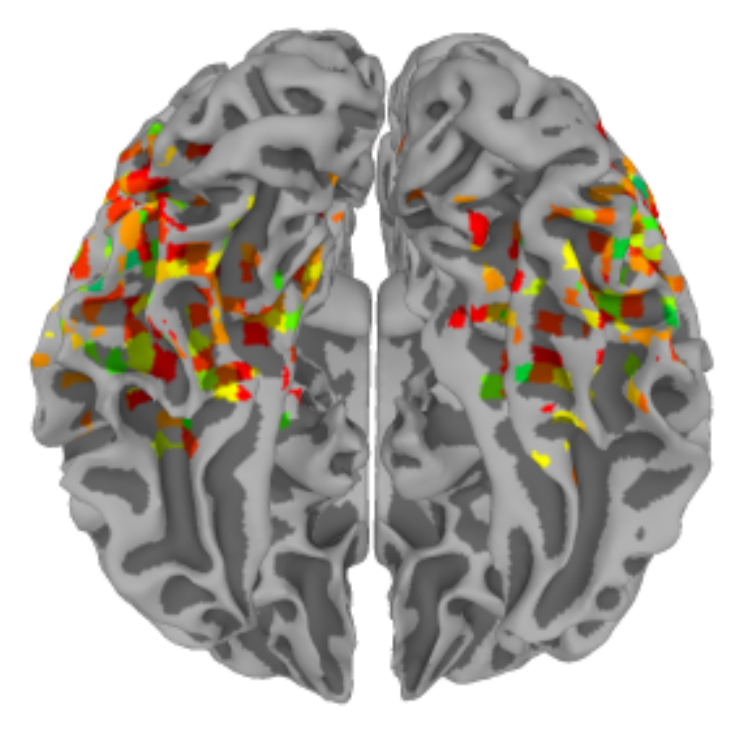}\\
			\centering (D) R105--Shoe
		\end{minipage}
		\begin{minipage}{0.04\linewidth}
			\includegraphics[width=0.4\textwidth]{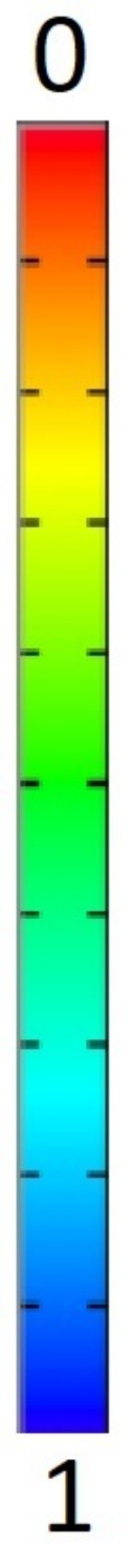}\\
		\end{minipage}
		\begin{minipage}{0.22\linewidth}
			\includegraphics[width=0.9\textwidth]{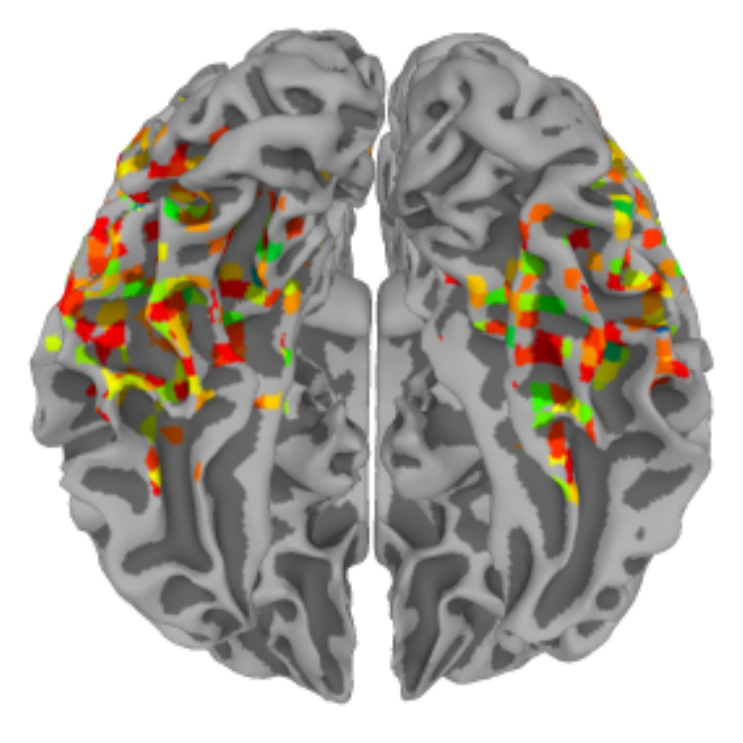}\\
			\centering (E) R105--House
		\end{minipage}
		\begin{minipage}{0.22\linewidth}
			\includegraphics[width=0.9\textwidth]{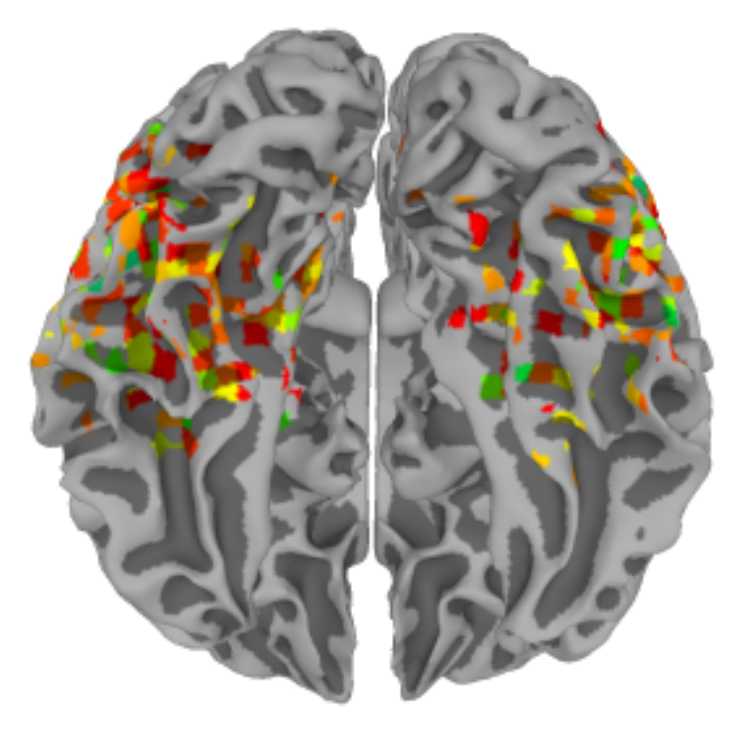}\\
			\centering (F) R105--Scramble
		\end{minipage}
		\begin{minipage}{0.22\linewidth}
			\includegraphics[width=0.9\textwidth]{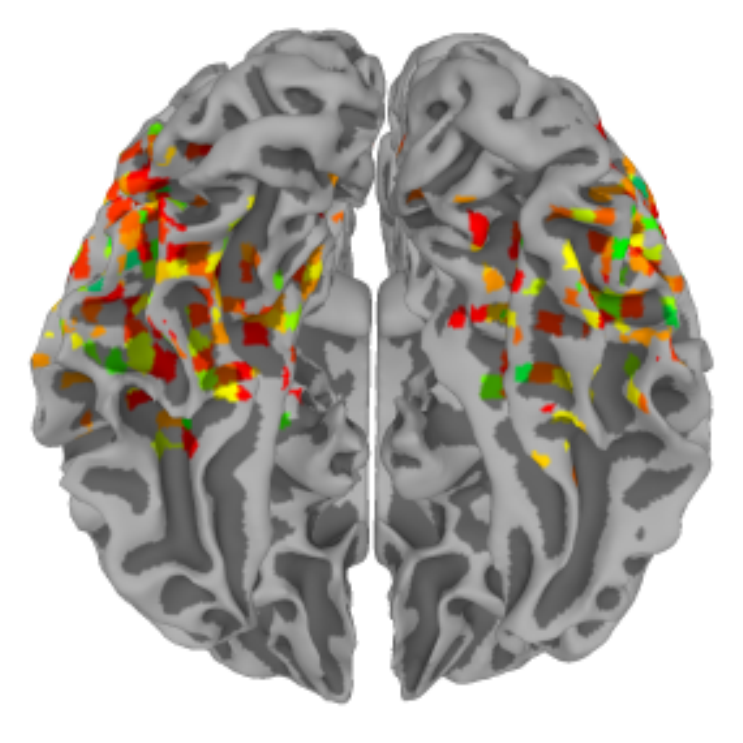}\\
			\centering (G) R105--Bottle
		\end{minipage}
		\begin{minipage}{0.22\linewidth}
			\includegraphics[width=0.9\textwidth]{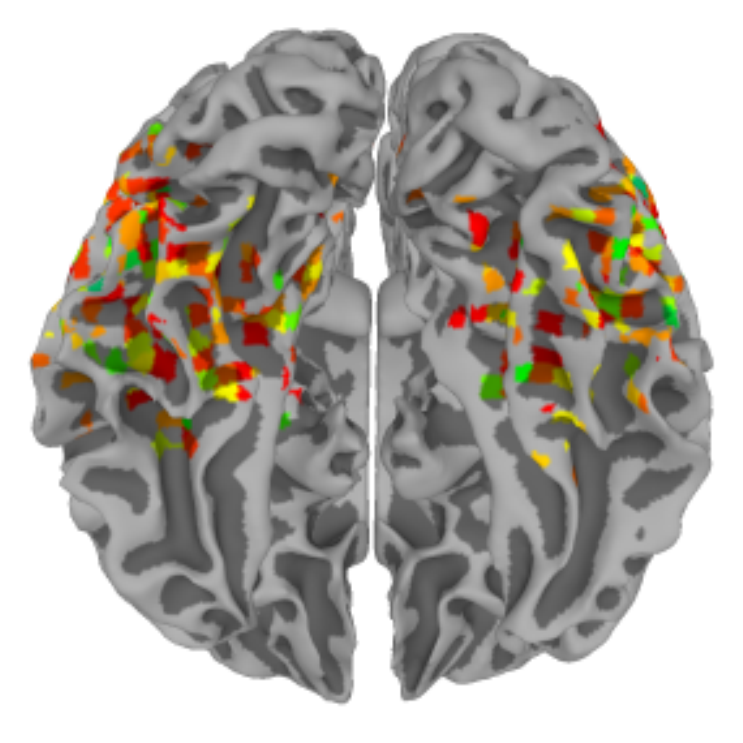}\\
			\centering (H) R105--Chair
		\end{minipage}
		\begin{minipage}{0.04\linewidth}
			\includegraphics[width=0.4\textwidth]{Fig5bar}\\
		\end{minipage}		
		\begin{minipage}{0.22\linewidth}
			\includegraphics[width=0.9\textwidth]{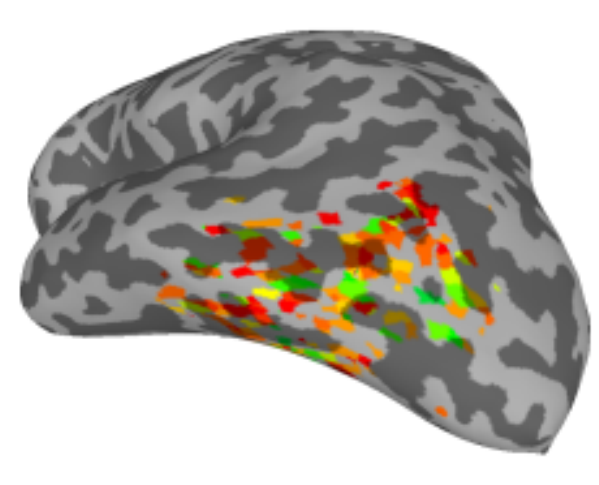}\\
			\centering (I) R107--Word
		\end{minipage}
		\begin{minipage}{0.22\linewidth}
			\includegraphics[width=0.9\textwidth]{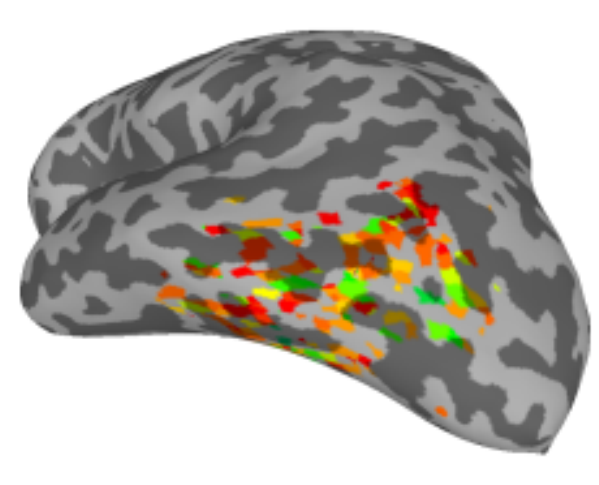}\\
			\centering (J) R107--Object
		\end{minipage}
		\begin{minipage}{0.22\linewidth}
			\includegraphics[width=0.9\textwidth]{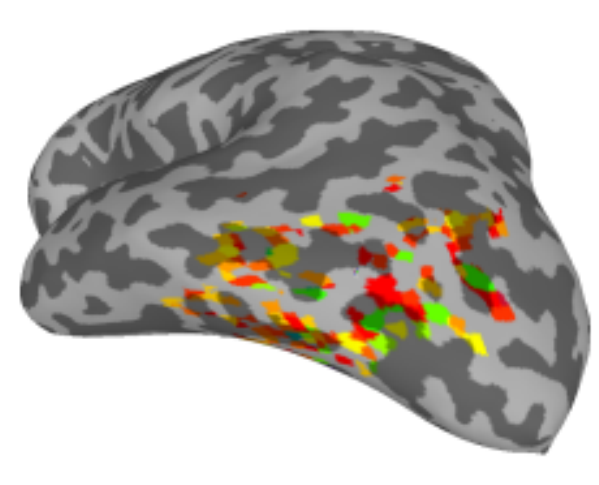}\\
			\centering (K) R107--Consonant
		\end{minipage}
		\begin{minipage}{0.22\linewidth}
			\includegraphics[width=0.9\textwidth]{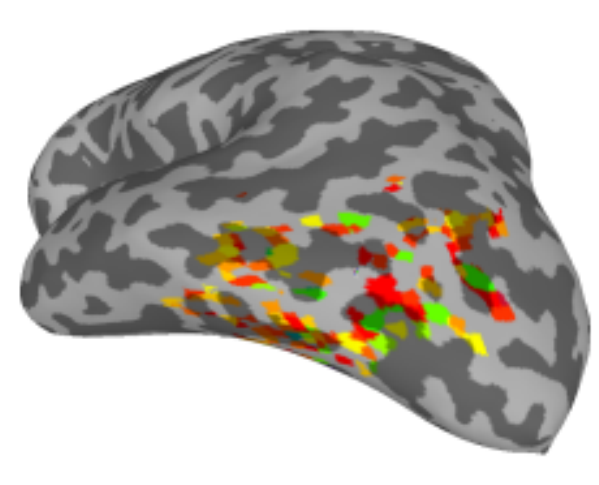}\\
			\centering (L) R107--Scramble
		\end{minipage}
		\begin{minipage}{0.04\linewidth}
			\includegraphics[width=0.4\textwidth]{Fig5bar}\\
		\end{minipage}	
		\caption{Representing neural activities in the linear embedded space by using R105 and R107 datasets}
		\label{fig:DRSANeuralActivities}
	\end{center}
	\vskip -0.2in
\end{figure*}

\subsection{Runtime analysis}
This section analyzes the runtime of the proposed method and other similarity methods by employing ROI-based datasets. As mentioned before, all results in this paper are generated by using a PC with certain specifications. Further, the runtime of DRSL is evaluated by using both hardware, i.e., CPU (DRSL) and GPU (DRSLg). Figure \ref{fig:Runtime} demonstrates the runtime of the mentioned methods, where runtime of other algorithms are scaled based on DRSL. In other words, the runtime of the proposed method is considered as a unit. As illustrated in this figure, BRSA and PCM generated the worse runtime because they must estimate a wide range of hyper-parameters for high-dimensional datasets. Further, the runtime of DRSL is similar to the regularized methods (LASSO and RSL), while those algorithms did not utilize any transformation function. Since GRSA and LRSL (same as DRSL) employ a min-batch of time-points, they produce better runtime in comparison with the regularized methods. By considering the performance of the proposed method in the previous section, DRSL generates acceptable runtime. As mentioned before, the proposed method utilizes gradient-based approaches that can rapidly reduce the time complexity of the optimization procedure. It is worth noting that runtime of the whole-brain datasets has the same tendency.

\subsection{DRSL iteration analysis}
In this section, we analyze the performance of DRSL by using different numbers of iterations. As the first step, we must calculate the Mean of Square Error ($MSE$) for DRSL technique, where better models generate lower error --- i.e., $MSE = \frac{1}{TSV}\sum_{\ell=1}^{S}\sum_{i=1}^{T}\sum_{j=1}^{V} \Big({x}_{ij}^{(\ell)} - \sum_{k=1}^{P} {d}_{ik}^{(\ell)}{\beta}_{kj}^{(\ell)}\Big)^2$. Figure \ref{fig:Iter} illustrates the performance of DRSL by using MSE, where different number of iterations are used --- i.e., $100, 500, 900, 1000, 1500, 2000, 2500, \text{ and } 3000$ iterations. Here, the network structure for DRSL is considered fixed and same as the previous section. As Figure \ref{fig:Iter} depicted, MSE in most of the datasets is reduced between $100$ iterations to $1000$ iterations. However, MSE does not have significant fluctuation in most of the datasets (except W231 dataset) after $1000$ iterations. Another advantage of using DRSL with $1000$ iterations in this paper was the trade-off between runtime and the performance --- please refer to~Figure~\ref{fig:Runtime}, Table~\ref{tbl:Cor}, and Table~\ref{tbl:Cls}. Based on these results, we have used $10 \times 100 = 1000$ iterations as the default value to generate all of the results in the previous sections.

\subsection{Analyzing the proposed regularization function}
In this section, we present some insights into how the proposed regularization function works. As the primary advantage, the regularization term $r(\mathbf{B}) = \sum_{j=1}^{V}\sum_{k=1}^{P}\alpha\big|\beta_{kj}\big| + 10\alpha(\beta_{kj})^{2}$ is convex, and it is consequently suitable for optimization.

As Figure \ref{fig:RT} depicted, the coefficient $\alpha=10$ can increase the sensitivity of the regularization term to handle small change of $\beta$ values in comparison with the coefficient $\alpha=1$. There are two advantages to use $\alpha=10$. First, since the neural activities are highly correlated, the difference of $\beta$ values in different categories of stimuli is indeed small. Second, it can remove the effect of random noises on the performance of analysis --- specifically on a dataset with a low rate of SNR.

\subsection{Representing neural signatures}
In this section, we visualize some samples of neural signatures that are produced by DRSL --- based on $\mathbf{\widetilde{B}}$ matrix --- for datasets R105 and R107. Figure \ref{fig:DRSANeuralActivities} shows these generated neural signatures. Here, we demonstrate the left side of the brain for R107 because there is no ROI region on the right side for this dataset \cite{Khaligh14}. As depicted in this figure, while some of the neural activities are significantly distinctive across categories of stimuli, the rest of them are highly correlated. For instance, word stimuli and object category have a strong correlation as the meaningful concepts in R107, whereas the neural signatures of these categories are completely different with non-meaningful concepts, i.e., consonant and scramble stimuli. Similarly, the non-meaningful concepts also have a strong correlation in this dataset. As another alternative, we can compare the meaningful concepts in R105 with the neural signature generated for scramble stimuli.

\section{Discussion and Conclusion}
Similarity analysis is one of the crucial steps in most fMRI studies. The existing efforts on developing similarity approaches to analyze neural responses show some promising results. However, there are still several long-standing challenges. First, classic similarity methods such as RSA (with OLS and GLM solutions) may not provide stable analysis --- especially when the covariance of the neural activities has a low rate of SNR. The proposed method employed a new regularization term that steeply penalized neural responses to handle the SNR issue. Second, most regularized similarity approaches such as LASSO, RSL, and GRSA use a linear method for generating the neural signatures. In this paper, we used a parametric, nonlinear transformation function to map the neural responses to a lower-dimensional latent space. The proposed method enabled us to discover more complex feature representations for similarity analysis. Third, most nonlinear similarity approaches such as BRSA and PCM cannot provide a time-efficient analysis on data with a large number of subjects, a broad ROI, or even whole-brain fMRI images. Alternatively, the proposed method utilized a gradient-based optimization technique that rapidly reduced the runtime for analyzing high-dimensional fMRI datasets.

The main difference between the proposed deep model and a regular network model lies in its application. A regular deep neural network is mostly used in the form of a supervised structure --- i.e., it maps input space to supervised labels. However, our method employed the unsupervised deep neural network (i.e., multiple stacked layers of nonlinear transformations) as the parametric kernel function that maps the voxel space to a lower-dimension, information-rich feature space. In some sense, the transformation function employed in DRSL is similar to the one utilized in the Deep Canonical Correlation Analysis (DCCA) for multi-view representational learning or Deep Hyperalignment (DHA) for aligning neural activities of multi-subject data across subjects. Nevertheless, DCCA and DHA cannot be used for similarity analysis. The primary objective functions in these approaches map neural activities of each subject to a shared space --- where the transformed features can improve the performance of a classification model. Unlike DCCA and DHA, our method is formulated as a nonlinear multi-set regression problem --- where it must find better regressors to map the matrix of neural responses and the design matrix.

In summary, this paper introduces a deep learning approach for Representational Similarity Analysis (RSA) in order to provide accurate similarity (or distance) analysis in multi-subject fMRI data. Deep Representational Similarity Learning (DRSL) can handle fMRI datasets with noise, sparsity, nonlinearity, high-dimensionality (broad ROI or whole-brain data), and a large number of subjects. DRSL utilizes gradient-based optimization approaches and generates an efficient runtime on large datasets. Further, DRSL is not limited by a restricted fixed representational space because the transformation function in DRSL is a multi-layer neural network, which can separately implement any nonlinear function for each subject to transfer the neural activities into an embedded linear space. To evaluate the performance of the proposed method, multi-subject fMRI datasets with various tasks --- including visual stimuli, decision making, flavor, and working memory --- are employed for running the empirical studies. The results confirm that DRSL achieves superior performance to other state-of-the-art RSA algorithms for evaluating the similarities between distinctive cognitive tasks. In the future, we will plan to use the proposed method for improving the performance of other techniques in fMRI analysis, e.g., multi-modality, and hub detection.

\section{Proof of Lemma \ref{lm:GradientB}}
\vspace{1mm}
\underline{Step 1}:\\

Suppose $\mathbf{X}$ and $\mathbf{Y}$ are matrices of any size. And there is a differentiable mapping such that $\mathbf{Y}=f_1(\mathbf{X})$. If $f_2$ is a real-valued function differentiable with respect to $\mathbf{Y}$, then we have:
\begin{equation*}
	\begin{gathered}
		\cfrac{\partial f_2(\mathbf{Y})}{X_{ij}} = \mathrm{tr}\left[ \cfrac{\partial \mathbf{Y}}{\partial X_{ij}}\,\left( \cfrac{\partial f_2(\mathbf{Y})}{\partial \mathbf{Y}} \right)^{\top} \right].
	\end{gathered}
\end{equation*}
A simple application of the chain rule gives:

\begin{gather*}
\begin{split}
\cfrac{\partial f_2(\mathbf{Y})}{\partial X_{ij}} & = \sum_{k}\sum_{l}\cfrac{\partial f_2(\mathbf{Y})}{\partial Y_{kl}}\,\cfrac{\partial Y_{kl}}{\partial X_{ij}}\\
& = \sum_{k}\left( \cfrac{\partial \mathbf{Y}}{\partial X_{ij}} \right)_{k.}\,\left( \cfrac{\partial f_2(\mathbf{Y})}{\partial \mathbf{Y}} \right)^{\top}_{.k}\\
& = \mathrm{tr}\left[ \cfrac{\partial \mathbf{Y}}{\partial X_{ij}}\,\left( \cfrac{\partial f_2(\mathbf{Y})}{\partial \mathbf{Y}} \right)^{\top} \right]. 
\end{split}
\end{gather*}
\vspace{1mm}
\underline{Step 2}:\\

We can proof the following equality for any matrix $\bf A\text{, }X\text{, }B$ in $\mathbb{R}$ space:
\begin{gather*} \bf
	\cfrac{\partial \| AX+B \|_{\mathnormal{F}}^{\mathnormal{2}}}{\partial  X} = 2A^{\top}(AX+B).
\end{gather*}
Let $ \mathbf{D}^{(ij)} $ denote $ \cfrac{\partial \bf AX}{\partial X_{ij}} $ and $ \mathbf{Y} $ denote $ \mathbf{AX+B} $.

Firstly, we have
\begin{gather*}
D^{(ij)}_{kl} = \cfrac{\partial (AX)_{kl}}{\partial X_{ij}} = \cfrac{\partial \sum_{m}A_{km}X_{ml}}{\partial X_{ij}} = \begin{cases}
0, \ l \not= j\\
A_{ki}, \ \text{ otherwise }
\end{cases}
\end{gather*}
According to Step 1, we have
\begin{gather*}
\begin{split}
\cfrac{\partial \| \mathbf{AX+B} \|_{F}^{2}}{\partial X_{ij}} & = \mathrm{tr}\left[ \cfrac{\partial (\mathbf{AX+B})}{\partial X_{ij}}\,\left( \cfrac{\partial \| \mathbf{AX+B} \|_{F}^{2}}{\partial (\mathbf{AX+B})} \right)^{\top} \right]\\
& = 2\mathrm{tr}\left( \mathbf{D}^{(ij)}Y^{\top} \right)\\
& = 2\sum_{k}\sum_{l}D^{(ij)}_{kl}Y_{kl}\\
& = 2\sum_{k} D^{(ij)}_{kj}Y_{kj}\\
& = 2\sum_{k}A_{ki}Y_{kj}\\
& = 2(A^{\top}Y)_{ij}. 
\end{split}
\end{gather*}
\vspace{1mm}
\underline{Step 3}:\\

By considering Steps 1 and 2, we have:

\begin{equation*}
\begin{gathered}
\nabla_{\mathbf{B}^{(\ell)}}(J_{R}^{(k,\ell)})\\ = \sum_{i\in \Psi^{(k,\ell)}}\cfrac{\partial }{\partial \mathbf{B}^{(\ell)}}\| f(\mathbf{x}_{i.}^{(\ell)};\theta^{(\ell)}) - \mathbf{d}_{i.}^{(\ell)}\mathbf{B}^{(\ell)} \|_{2}^{2} + \cfrac{\partial }{\partial \mathbf{B}^{(\ell)}}r(\mathbf{B}^{(\ell)})\\
= -2\sum_{i\in\Psi^{(k,\ell)}}\big(\mathbf{d}_{i.}^{(\ell)} \big)^{\top}\left( f(\mathbf{x}_{i.}^{(\ell)};\theta^{(\ell)}) - \mathbf{d}_{i.}^{(\ell)}\mathbf{B}^{(\ell)} \right) + \cfrac{\partial }{\partial \mathbf{B}^{(\ell)}}r(\mathbf{B}^{(\ell)})
\end{gathered}
\end{equation*}
For the last term $\cfrac{\partial }{\partial \mathbf{B}^{(\ell)}}r(\mathbf{B}^{(\ell)})$, we have:
\begin{equation*}
\begin{gathered}
\cfrac{\partial }{\partial \beta_{ab}^{(\ell)}}r(\mathbf{B}^{(\ell)}) =  \sum_{j=1}^{V}\sum_{k=1}^{P}\alpha\cfrac{\partial \big| \beta_{kj}^{(\ell)} \big|}{\partial \beta_{ab}^{(\ell)}} + 10\alpha\cfrac{\partial (\beta_{kj}^{(\ell)})^{2}}{\partial \beta_{ab}^{(\ell)}} \\
 = \alpha\cfrac{\beta_{ab}^{(\ell)}}{\big| \beta_{ab}^{(\ell)} \big|} + 20\alpha\beta_{ab}^{(\ell)}\\
 = (\alpha\mathrm{sign}(\mathbf{B}^{(\ell)}) + 20\alpha\mathbf{B}^{(\ell)})_{ab}
\end{gathered}
\end{equation*}
Therefore, we obtain that $\cfrac{\partial }{\partial \mathbf{B}^{(\ell)}}r(\mathbf{B}^{(\ell)}) = \alpha\mathrm{sign}(\mathbf{B}^{(\ell)}) + 20\alpha\mathbf{B}^{(\ell)}$, completing the proof. $\qquad \blacksquare$

\section*{Information Sharing Statement}
In this paper we use the publicly available Open Neuro datasets that can be found via the site link: \url{https://openneuro.org/}. Further, we have shared a preprocessed version of datasets in MATLAB format at \url{https://easydata.learningbymachine.com/}. In addition, the proposed method can be accessed by using our GUI-based toolbox --- i.e., available at \url{https://easyfmri.learningbymachine.com/}.

\section*{Compliance with Ethical Standards}
\section*{Conflict of Interests }
All authors declare that they have no conflict of interest.
\section*{Ethical Approval}
This article does not contain any studies with human participants or animals performed by any of the authors.
\begin{acknowledgements}
This work was supported by the National Natural Science Foundation of China (Nos. 61876082, 61732006, 61861130366), the National Key R\&D Program of China (Grant Nos. 2018YFC2001600, 2018YFC2001602, 2018ZX10201002), the Research Fund for International Young Scientists (NSFC Grant No. 62050410348), the Royal Society-Academy of Medical Sciences Newton Advanced Fellowship (No. NAF$\backslash$R1$\backslash$180371), and the Alberta Machine Intelligence Institute (Amii). This paper is in honor of our friends Pouneh Gorji, Arash Pourzarabi, and all other victims of the Ukraine Flight PS752.
\end{acknowledgements}


%
%

\bibliographystyle{spbasic}      
\bibliography{DRSL}   

%
%

\end{document}